\documentclass[12pt,a4paper]{article}
\usepackage[utf8]{inputenc}
\usepackage{natbib}

\usepackage[top=35truemm,bottom=35truemm,left=27truemm,right=27truemm]{geometry}

\usepackage{setspace}
\usepackage{amsthm, amssymb, amsmath, appendix}
\usepackage{mathtools}
\mathtoolsset{showonlyrefs=true}

\theoremstyle{definition}
\newtheorem{theorem}{Theorem}
\newtheorem{lem}{Lemma}

\newtheorem{prop}{Proposition}

\newtheorem{observation}{Observation}
\newtheorem{remark}{Remark}

\allowdisplaybreaks

\usepackage{tikz}
\usetikzlibrary{intersections, calc, arrows.meta}

\makeatletter

\@addtoreset{figure}{section}
\@addtoreset{table}{section}
\makeatother 

\title{The Cesàro average criterion on infinite utility streams and its extensions\thanks{The author is grateful to the two anonymous referees for their helpful comments and suggestions.
The author also thanks Kohei Kamaga, Noriaki Kiguchi, Kaname Miyagishima, Nozomu Muto, Marcus Pivato, Koichi Tadenuma, Norio Takeoka, Tsubasa Yamashita, Shohei Yanagita, and Hide-fumi Yokoo for their helpful comments.
The previous title of this paper was ``Impartial utilitarianism on infinite utility streams.''
This research did not receive any specific grant from funding agencies in the public, commercial, or not-for-profit sectors.}}
\author{Kensei Nakamura\thanks{Graduate School of Economics, Hitotsubashi University, Kunitachi, Tokyo 186-8601, Japan. E-mail : kensei.nakamura.econ@gmail.com}}
\date{\today}

\begin{document}

\maketitle

\vspace{3mm}

\begin{abstract}
    When evaluating policies that affect future generations, the most commonly used criterion is the discounted utilitarian rule. 
    However, in terms of intergenerational fairness, it is difficult to justify prioritizing the current generation over future generations. 
    This paper axiomatically examines impartial utilitarian rules over infinite-dimensional utility streams. 
    We provide simple characterizations of the social welfare ordering that evaluates utility streams by their long-run average on the domain where the average exists.
    Furthermore, we derive the necessary and sufficient conditions for the same axioms to hold in a more general domain, the set of bounded utility streams. 
    Some of these results are closely related to the Banach limits, a well-known generalization of the classical limit concept for streams. 
    Thus, this paper can be seen as proposing an appealing subclass of Banach limits through axiomatic analysis. 
    

\end{abstract}

\newpage
\section{Introduction}
\label{sec_intro}

Many economic policies have long-term effects, impacting future generations either positively or negatively. For instance, when addressing climate change, neglecting the issue and maintaining the status quo can cause severe harm to future generations. In contrast, regulating economic activity might improve the living standards of distant future generations, but this often comes at the expense of the current generation. These policies should be evaluated and chosen with consideration for the conflicts among generations. Therefore, it is crucial to investigate acceptable criteria for these long-term policies.

The most widely used criteria for long-term policies are the discounted utilitarian rules.  
These rules exponentially discount the utility levels of future generations with a discount factor $\delta\in (0,1)$ and evaluate utility streams by summing them up. 
Despite their popularity, it has been pointed out that these rules excessively disregard future generations and are undesirable in terms of intergenerational fairness even when $\delta$ is close to $1$. 
For instance, as discussed in Chapter 2.A of \citet{stern2007economics}, if we exponentially discount the utility levels by 1\%  each period, then the utility of people 100 periods later is only about 37\% of the actual value. 
\citet{ramsey1928mathematical} stated that ``we do not discount later enjoyments in comparison with earlier ones, a practice which is ethically indefensible
and arises merely from the weakness of the imagination."
Following this point, this paper investigates utilitarian criteria that treat all generations equally. 
More specifically, we axiomatically examine the welfare criteria that evaluate bounded utility streams $(u_1, u_2, u_3, \cdots) \in \mathbb{R}^\mathbb{N}$ by their long-run average, or the Cesàro average, 
\begin{equation*}
    \lim_{T\rightarrow \infty} {1\over T} \sum_{t = 1}^T u_t 
\end{equation*}
if the limit exists.
We call these criteria the \textit{Cesàro average social welfare orderings}.

Our first main result (Theorem \ref{thm_chara_lCes}) provides two characterizations of the Cesàro average social welfare ordering on a \textit{restricted domain}---the set of bounded utility streams whose Cesàro average exists. 
In our result, weak versions of additive independence play a crucial role, instead of the axiom of separability, i.e., independence of unconcerned generations (cf.~\citealp{P2022JET,li2024simple}). Additive independence requires that the ranking between two streams remain unchanged if a common vector is added to both. Our weak versions postulate this invariance property when (i) the utility level of only one generation changes or (ii) streams are periodic, respectively. Together with basic axioms and novel consistency axioms regarding the time horizon, we show that each additivity axiom characterizes the Cesàro average social welfare ordering in the restricted domain.

In \citeauthor{D1965Econometrica}'s (\citeyear{D1965Econometrica}) seminal paper,  it was proven that there is no continuous social welfare ordering that satisfies the standard Paretian condition and impartiality between any two generations. 
Similar impossibility results have been established by \citet{BM2003Econometrica}, \citet{FM2003JME}, and others. 
Under these impossibilities, many utilitarian social welfare criteria have been proposed and axiomatized, such as the catching-up criteria, the overtaking criteria (\citealp{atsumi1965neoclassical,von1965existence}), the dominance-in-tail criteria (\citealp{BM2007Book}), and their fixed-step versions (\citealp{lauwers1997infinite,FM2003JME,KK2009SCW}). These criteria can be obtained by sacrificing continuity or completeness. 
However, considering applications in economic analysis, these properties are not merely technical but are also normatively appealing and possibly essential.  
Instead of giving up on them, we escape these impossibilities by slightly weakening the Paretian principle. 

Next, we address the problem of the restricted domain: Because not all bounded utility streams have a Cesàro average, we cannot use the Cesàro average social welfare orderings to evaluate these streams. Although \citet{P2022JET} and \citet{li2024simple} provided  characterizations of the Cesàro average social welfare function in essentially similar domains, as pointed out by \citet{pivato2024intergenerational}, a shortcoming of their characterizations is that the set of streams where the Cesàro average can be defined is somewhat restricted. 
Similarly, our characterization results in the restricted domain face the problem of domain restriction.
To address this issue, we examine the implications of the axioms in the first result within a more general domain, the set of bounded utility streams. 
We show that social welfare orderings satisfying these axioms are compatible with long-run total utility criteria, such as the catching-up criterion or its fixed-step version, and that acceptable rankings in the larger domain depend on which set of axioms is adopted (Theorems \ref{thm_genCes1} and \ref{thm_genCesfixed}). 

Furthermore, we investigate fully linear extensions of the Cesàro average social welfare ordering to the general domain. In finite-population social welfare orderings, the axiom of additive independence is often interpreted as indicating the degree of interpersonal comparability (\citealp{dAspremont1977equity,roberts1980interpersonal,BBD2002BOOK}). 
When imposing an extension of additive independence under this interpretation, it is quite natural to consider fully additive independence. 
We show that if we extend the Cesàro average social welfare ordering on the restricted domain to the larger domain with full additivity, then these rankings are represented by linear social welfare functions characterized by simple inequalities (Theorems \ref{thm_full1} and \ref{thm_full2}).
These results provide upper and lower bounds for the evaluations of each utility stream. These bounds are related to the catching-up criterion or its fixed-step version, respectively. Moreover, these functions are special cases of the \textit{Banach limits}, a well-known generalization of the classical limit concept for streams. 
Thus, this paper can be considered as proposing appealing subclasses of Banach limits through axiomatic analysis.

This paper is organized as follows: 
Section \ref{subsec_literature} discusses the  related literature. 
Section \ref{sec_def} introduces utility streams and social welfare orderings. 
Section \ref{sec_propCes} examines properties of the  Cesàro average and the restricted domain. 
Section \ref{sec_axiom} considers desirable properties for social welfare orderings and formalizes them as axioms. 
The main results of this paper are presented in Section \ref{sec_main}. 
Section \ref{sec_chara_rest} provides our first characterizations of the Cesàro social welfare ordering in the restricted domain. 
Section \ref{sec_genCes} studies the implications of the axioms of the first result on the larger domain, and Section \ref{sec_fulladd} explores fully additive extensions of the Cesàro social welfare orderings. 
Section \ref{sec_conc} provides concluding remarks. 
In Appendix, we prove some of the results and discuss the independence of the axioms and the existence of the social welfare orderings that we characterize. 

\subsection{Related Literature}
\label{subsec_literature}

We discuss the literature related to the Cesàro average social welfare functions. 
\citet{P2022JET} and \citet{li2024simple} considered preferences over streams of objects and characterized the Cesàro average functions associated with instantaneous utility functions. 
The key axiom in these papers is the axiom of separability. 
In their abstract setups, our key axioms of additivity cannot be defined in a natural way since the additive operator over objects is not defined. 
By directly examining the utility stream, we provide another foundation for the Cesàro average social welfare functions and examine their properties in the larger domain. 
The novel consistency axioms also play a key role in our theorems. 
\citet{marinacci1998axiomatic} characterized similar classes of preferences over streams of lotteries. 

\citet{lauwers1995time,lauwers1998intertemporal} examined linear functions over bounded utility streams and characterized the class of Cesàro average functions we have obtained in Theorem \ref{thm_full1}. 
They considered a stronger impartiality axiom requiring that for each bijection $\pi: \mathbb{N} \to\mathbb{N}$, if $\lim_{t\rightarrow \infty} \pi (t) / t = 1$, then permuting generations by $\pi$ should not affect social welfare. 
Lauwers provided an axiomatic foundation using this impartiality axiom and a Paretian axiom, given the linearity of welfare functions. 
Compared with this result, Theorem \ref{thm_full1} in our paper provides an if-and-only-if foundation for the same class of social welfare orderings, using the simple impartiality axiom for finite permutations. 
Furthermore, while the proof of \citet{lauwers1995time,lauwers1998intertemporal} relies on results known in functional analysis, our proof is elementary.

\citet{JV2018TE}  characterized the social welfare relations represented as the limit of discounted utilitarianism as the discount factor approaches $1$ (note that they are closely related to the Cesàro average social welfare orderings, as shown in Observation \ref{obs_duCes} in this paper).
As in Theorems \ref{thm_full1} and \ref{thm_full2} of our paper, they used the axiom of full additivity to characterize the social welfare relations. 
The primary distinction lies in the Paretian conditions: Compared to our result, they obtained the characterization results, using a stronger Paretian principle and giving up on completeness and continuity. 

\citet{jonsson2015utilitarianism} partially characterized the strict part of the Cesàro average functions. They showed that if a social welfare ordering satisfies several properties, then for all pairs of utility streams, if the Cesàro average of one is greater than that of the other, then the former is strictly better than the latter. 
They did not examine the converse, the cases where the Cesàro averages are equal or their Cesàro averages do not exist. 
Our paper addresses these remaining problems. 
\citet{khan2018planning} also examined social welfare orderings respecting criteria similar to the catching-up criterion. 
\citet{asheim2025exploring} considered fully anonymous utilitarian rules by sacrificing Paretian conditions.

In a more general setting, \citet{pivato2014additive} examined the functions with anonymous additive representations. \citet{pivato2023cesaro} considered an infinite population of individuals dispersed throughout time and space, and characterized Cesàro average social welfare orderings using a separability axiom. 

\section{Definitions and Notations}
\label{sec_def}
This section presents several definitions and notations for utility streams and social welfare orderings. 

Let $\mathbb{R}^\mathbb{N}$ denote the set of utility streams. 
A typical element is denoted by $ \mathbf{u} = (u_1, u_2,  \cdots, u_t, \cdots ) \in \mathbb{R}^\mathbb{N}$, where $u_t$  is the well-being of  generation $t$. 
For all $T \in \mathbb{N}$, 
let $ \mathbf{u}_{[1,\cdots, T]} = (u_1, u_2, \cdots, u_T)$ and $\mathbf{u}_{[T+ 1, \cdots, \infty)} = (u_{T+1}, u_{T+2}, \cdots )$ denote the $T$-head and the $T$-tail of $  \mathbf{u}$ respectively. 
For all subsets $M\subset \mathbb{N}$, let $\mathbf{1}_M$ be a utility stream $\mathbf{v}$ such  that $v_t = 1$ for $t\in M$ and $v_t = 0$ otherwise.
For example, $\mathbf{1}_\mathbb{N} = (1, 1,1,\cdots )$. 
For any finite-dimensional vector $\mathbf{u}_{[1,\cdots, T]} \in \mathbb{R}^T$, let $[\mathbf{u}_{[1,\cdots, T]}]_\text{rep} = (u_1, \cdots , u_T, u_1, \cdots , u_T, u_1,  \cdots , u_T, \cdots ) \in \mathbb{R}^\mathbb{N}$. We call these streams \textit{periodic streams}.
For all $\mathbf{u}, \mathbf{v} \in \mathbb{R}^\mathbb{N}$, we write $ \mathbf{u} \geq \mathbf{v}$ if $u_t \geq v_t$ for all $t\in \mathbb{N}$. 

A \textit{permutation} is a bijection $\pi :\mathbb{N} \rightarrow \mathbb{N}$. 
For all $\mathbf{u} \in \mathcal{D}$, we write $ \mathbf{u}^\pi = (u_{\pi(1)}, u_{\pi(2)}, u_{\pi(3)}, \cdots ) $.
Let $\Pi$ denote the set of permutations.
We say that a permutation $\pi : \mathbb{N} \rightarrow \mathbb{N}$ is a \textit{finite permutation} if there exists  
$T \in \mathbb{N}$ such that for all $t \geq T$,  $\pi(t) = t$. 
Let $\Pi^{\text{fin}}$ denote the set of  finite permutations.

For all $\mathbf{u} \in \mathbb{R}^\mathbb{N}$ and $T\in\mathbb{N}$, denote the arithmetic mean of $\mathbf{u}_{[1,\cdots, T]}$ by $\mu_T (\mathbf{u})$, i.e., $\mu_T (\mathbf{u}) = \frac{1}{T} \sum_{t = 1}^{T} u_t$.
For all $\mathbf{u}\in \mathbb{R}^\mathbb{N}$ such that $\mu_T (\mathbf{u})$ converges to some number as $T\rightarrow \infty$, we define  the \textit{Cesàro average} $\mu_\infty (\mathbf{u})$ as follows:
\begin{equation*}
    \mu_\infty ( \mathbf{u}) = \lim_{T \rightarrow \infty } \mu_T ( \mathbf{u}) = \lim_{T \rightarrow \infty} {1 \over T} \sum_{t = 1 }^{T} u_t. 
\end{equation*}

Let $\mathcal{D}$ represent the generic domain of utility streams.  
This paper examines the following two domains. 
The first one is the set of all bounded utility streams, denoted by  $\ell^\infty = \{  \mathbf{u} \in \mathbb{R}^\mathbb{N} ~ | ~ \sup_{t \in \mathbb{N}} |u_t| < + \infty\}$.
This domain is one of the most standard domains in the literature.
The second one is a subset of $\ell^\infty$. 
Let $\ell^\text{Ces}$ denote the set of utility streams whose Cesàro average exists.
That is,  $\ell^\text{Ces} = \{ \mathbf{u} \in \ell^\infty ~ | ~ \text{there exists $\mu_\infty (\mathbf{u}) \in \mathbb{R} $}\} $. 
We assume that both $\ell^\infty$ and $\ell^\text{Ces}$ are endowed with the sup-norm topology. 
The next section examines the properties of the restricted domain and the Cesàro average. 

A binary relation $\succsim$ on $\mathcal{D}$  is a \textit{social welfare quasi-ordering} if it is reflexive and transitive.\footnote{
A binary relation $\succsim$ on $\mathcal{D}$  is reflexive if for all $\mathbf{u} \in \mathcal{D}$, $\mathbf{u} \succsim \mathbf{u}$; 
A binary relation $\succsim$ on $\mathcal{D}$  is transitive if for all $\mathbf{u}, \mathbf{v},  \mathbf{w} \in \mathcal{D}$, $ \mathbf{u} \succsim \mathbf{v}$ and $\mathbf{v} \succsim \mathbf{w}$ imply $\mathbf{u} \succsim \mathbf{w}$. 
}  
A binary relation $\succsim$ is a \textit{social welfare ordering} if it is a complete social welfare quasi-ordering.\footnote{A binary relation $\succsim$ on $\mathcal{D}$  is \textit{complete} if for all $ \mathbf{u},  \mathbf{v} \in \mathcal{D}$, $ \mathbf{u} \succsim \mathbf{v} $ or $ \mathbf{v}  \succsim \mathbf{u} $. }
For all $ \mathbf{u}, \mathbf{v} \in \mathcal{D}$, we use $\mathbf{u} \succsim \mathbf{v}$ to indicate that $\mathbf{u}$ is judged to be  at least as good as $\mathbf{v}$. 
The asymmetric and symmetric parts of $\succsim$ are denoted by $\succ$ and $\sim$, respectively:
We write $\mathbf{u} \sim \mathbf{v}$ when the two streams $\mathbf{u}$ and $\mathbf{v}$ are considered socially indifferent; we write $ \mathbf{u} \succ \mathbf{v}$ when  $ \mathbf{u}$ is deemed socially better than $\mathbf{v}$.
We say that $\succsim$ on $\mathcal{D}$  is represented by  a \textit{social welfare function} $W: \mathcal{D} \rightarrow \mathbb{R}$ if for all $\mathbf{u},  \mathbf{v} \in \mathcal{D} $,
\begin{equation*}
    \mathbf{u} \succsim \mathbf{v}\iff W( \mathbf{u}) \geq W( \mathbf{v}) .
\end{equation*}
A function $W:\mathcal{D}\rightarrow \mathbb{R}$ \textit{respects} $\succsim$ if $\mathbf{u} \succsim \mathbf{v}$ implies $W( \mathbf{u}) \geq W( \mathbf{v})$ for all $\mathbf{u}, \mathbf{v}\in \mathcal{D}$. 

We say that a social welfare ordering $\succsim$ on $\mathcal{D}$ is represented by a \textit{Cesàro average (social welfare) function} $W : \mathcal{D}\rightarrow \mathbb{R}$ if $W (\mathbf{u}) = \mu_\infty (\mathbf{u})$ for all $\mathbf{u} \in \ell^\text{Ces}$. 
Note that if the domain is $\ell^\text{Ces}$, this function is uniquely determined. 
Also, we refer to social welfare orderings represented by a Cesàro average function as \textit{Cesàro average social welfare orderings}. 

Finally, we introduce several properties of functions on $\mathcal{D}$. 
We say that a function $W:\mathcal{D} \rightarrow \mathbb{R}$ is \textit{weakly monotone} if for all $\mathbf{u}, \mathbf{v} \in \mathcal{D}$, $\mathbf{u} \geq \mathbf{v} + \varepsilon \mathbf{1}_\mathbb{N}$ for some $\varepsilon > 0$ implies $W(\mathbf{u}) > W(\mathbf{v})$. 
We say that a function $W:\mathcal{D} \rightarrow \mathbb{R}$ is \textit{tail-monotone} if for all $  \mathbf{u},  \mathbf{v} \in \mathcal{D}$, $\mathbf{u}_{[T, \cdots, \infty)} \geq \mathbf{v}_{[T, \cdots, \infty)} + \varepsilon \mathbf{1}_\mathbb{N}$ for some $T\in\mathbb{N}$ and $\varepsilon > 0$ implies $W( \mathbf{u}) > W( \mathbf{v})$. 
Also, a function $W: \mathcal{D} \rightarrow \mathbb{R}$ is \textit{linear} if for all $\mathbf{u}, \mathbf{v} \in \mathcal{D}$ and $\alpha, \beta \in \mathbb{R}$, $W(\alpha \mathbf{u} + \beta \mathbf{v}) = \alpha W(\mathbf{u}) + \beta W(\mathbf{v})$.

\section{Properties of the Cesàro Average}
\label{sec_propCes}

This section examines properties of the Cesàro average and the restricted domain $\ell^\text{Ces}$. All proofs of the results in this section are provided in Appendix.
First, we discuss how the Cesàro average is closely related to the discounted utilitarian rule. 
Given a discount factor $\delta \in (0, 1)$, the \textit{discounted utilitarian (social welfare) function} $\sigma_\delta : \mathcal{D} \rightarrow \mathbb{R}$ is defined as follows: For all $\mathbf{u} \in \mathcal{D}$, 
\begin{equation*}
    \sigma_\delta  (\mathbf{u}) = (1 - \delta) \sum_{t = 1}^{\infty} \delta^{t-1} u_t. 
\end{equation*}
These social welfare functions undervalue the welfare levels of future generations by the discount factor $\delta \in (0,1)$, thereby violating the principle of intergenerational fairness. 
As the discount factor $\delta$ approaches $1$, then $\sigma_\delta$ treats each generation more equally.
Intuitively, the limit of the discounted utilitarian function $\sigma_\delta$ and the Cesàro average  are very  similar in the sense that both  evaluate utility streams by the sum of utility levels of each generation and treat each generation equally.  
Indeed, the following statements can be established:

\begin{observation}
\label{obs_duCes}
    The following two statements hold: 
\begin{enumerate}
    \item For all $\mathbf{u} \in \mathbb{R}^\mathbb{N}$ and $k\in\mathbb{N}$, 
    \begin{equation*}
        \liminf_{T\rightarrow \infty} \mu_{kT} ( \mathbf{u}) \leq \liminf_{\delta \rightarrow 1^-}  \sigma_\delta (\mathbf{u}) \leq  \limsup_{\delta \rightarrow 1^-}  \sigma_\delta (\mathbf{u})  \leq  \limsup_{T\rightarrow \infty} \mu_{kT} ( \mathbf{u}).
    \end{equation*}
    \item For all  $\mathbf{u} \in \ell^\text{Ces}$, $ \mu_{\infty} (\mathbf{u}) = \lim_{\delta \rightarrow 1^-} \sigma_\delta (\mathbf{u})$.
\end{enumerate}
\end{observation}

The first statement shows a general relationship between the limit of the mean of generations from $1$ to $kT$ as $T$ goes to infinity and the limit of the sum of exponentially discounted utilities as the discount factor goes to $1$. Given a utility stream and a length $k$ of steps,  values that the discounted utilitarian rules can take in their limiting behavior are within the interval between the limit inferior and limit superior of the $k$-step average. 
Note that although the parameter $k$ does not play a crucial role, we include it here for later use. 
The second statement means that the Cesàro average $\mu_{\infty} ( \mathbf{u})$ can be interpreted as the limit of discounted utilitarian functions.
This result was also proved by \citet{F1880} directly. 
We obtain the same result as a corollary of the first statement. 

Next, we consider utility streams $\mathbf{u}$ such that the value $\lim_{t \rightarrow \infty} u_t$ exists. 
The Cesàro average has the following well-known properties:
\begin{observation}
\label{obs_limCes}
    For all $\mathbf{u} \in \mathbb{R}^\mathbb{N}$, if there exists  $\lim_{t\rightarrow \infty} u_t$, then $\mu_\infty ( \mathbf{u}) = \lim_{t\rightarrow \infty} u_t$. 
\end{observation}
This means that the Cesàro average is a generalization of the limit of utility streams. 

Finally, we examine the properties  of the  restricted domain $\ell^\text{Ces}$. 
The following statement holds:  

\begin{observation}
\label{obs_proLces}
    The set $\ell^\text{Ces}$ is a closed linear subspace of $\ell^\infty$. 
\end{observation}
    
This result means that the set $\ell^\text{Ces}$ is closed under element-wise addition, scalar multiplication, and the limit operation.
From this observation, we can see that, using the Hahn--Banach extension theorem, we can extend a linear function on $\ell^\text{Ces}$ to $\ell^\infty$ (Appendix \ref{app_exist}). 

\section{Axioms for Social Welfare Orderings}
\label{sec_axiom}

Many natural and reasonable properties for social welfare orderings, which we call \textit{axioms}, have been examined in the literature. 
We start with axioms of intergenerational fairness. 
The requirement for treating different generations equally is formalized using permutations. 
The first axiom is the most standard axiom of anonymity. 

\begin{description}
    \item[\bf Finite Anonymity.] For all $\mathbf{u} \in \mathcal{D}$ and  all $\pi \in \Pi^\text{fin}$, $\mathbf{u} \sim \mathbf{u}^\pi$. 
\end{description}

The second anonymity axiom focuses on periodic streams. 
This is the counterpart of the axiom called \textit{periodic fairness} in \citet{li2024simple}. 

\begin{description}
    \item[\bf Periodic Anonymity.] For all $\mathbf{u}\in \mathcal{D}$, all $T\in \mathbb{N}$, and all $\pi \in \Pi$, if $\{\pi(1), \pi (2), \cdots, \pi (T)\} = \{1, 2, \cdots, T \}$,
    then $[\mathbf{u}_{[1,\cdots, T]}]_\text{rep} \sim [\mathbf{u}^\pi_{[1,\cdots, T]}]_\text{rep}$. 
\end{description}

It is known that the standard axioms of efficiency and impartiality have severe tensions (\citealp{D1965Econometrica,BM2003Econometrica}). 
To escape from these impossibilities, we consider a weak axiom of efficiency.
The following axiom requires that if all generations prefer  $\mathbf{u}$ to  $\mathbf{v}$ and furthermore, the difference between them in each generation $t$ does not converge to zero as $t$ goes to infinity, then $\mathbf{u}$ should be socially better than $\mathbf{v}$. That is, if all generations consider  $\mathbf{u}$ to be sufficiently better than $\mathbf{v}$, then $\mathbf{u}$ is ranked to be strictly better than $\mathbf{v}$.\footnote{The same axiom was examined in \citet{M2015BER} and \citet{S2016SCW}.}
\begin{description}
    \item[\bf Uniform Pareto.] For all $\mathbf{u}, \mathbf{v} \in \mathcal{D}$, if $\mathbf{u} \geq \mathbf{v} + \varepsilon \mathbf{1}_\mathbb{N}$ for some $\varepsilon > 0$, then $\mathbf{u}\succ \mathbf{v}$.
\end{description}

The next axiom concerns the continuity of rankings. 
It postulates that social welfare evaluations should be robust to small changes in utility levels of each generation. 
\begin{description}
    \item[\bf Continuity.] For all $\mathbf{u}, \mathbf{v} \in \mathcal{D}$ and all sequences $\{\mathbf{u}^{k} \}_{k \in \mathbb{N}}$ in $\mathcal{D}$ such that $\mathbf{u}^k \rightarrow \mathbf{u}$ as $k \rightarrow \infty$ in the sup-norm topology, if $\mathbf{u}^{k} \succsim  \mathbf{v}$ for each $k \in \mathbb{N}$, then $ \mathbf{u} \succsim  \mathbf{v}$; if $ \mathbf{v} \succsim \mathbf{u}^{k}$ for each $k \in \mathbb{N}$, then $\mathbf{v} \succsim  \mathbf{u}$.  
\end{description}
Note that since $\ell^\infty$ and $\ell^\text{Ces}$ are closed sets (Observation \ref{obs_proLces}), regardless of the domain, we can apply this axiom to all sequences that converge to some point in  $\ell^\infty$. 
Also note that together with \textit{continuity},  \textit{uniform Pareto} implies that for all $\mathbf{u}, \mathbf{v}\in\mathcal{D}$, if $\mathbf{u}\geq \mathbf{v}$, then $\mathbf{u}\succsim \mathbf{v}$.\footnote{We prove this property. 
For all $ \mathbf{u}, \mathbf{v} \in \mathcal{D}$ with $\mathbf{u} \geq  \mathbf{v}$, \textit{uniform Pareto} implies that for all $k \in \mathbb{N}$, $\mathbf{u} + (1/k)\mathbf{1}_\mathbb{N} \succ \mathbf{v}$. 
As $k$ goes to infinity, the left-hand side converges to $\mathbf{u}$. 
By \textit{continuity}, we have $ \mathbf{u} \succsim \mathbf{v}$.  } 
Thus, under these two axioms, it does not happen that all generations weakly prefer $\mathbf{u}$ to $\mathbf{v}$ but  $\mathbf{u}$ is socially worse than $\mathbf{v}$. 

The following condition and its variants have been widely examined in the literature on finite or infinite dimensional social welfare orderings.\footnote{For other variants in infinite dimensional social welfare orderings, see \citet{AT2004ET}, \citet{banerjee2006extension}, and \citet{BM2007JET}.} 
\begin{description}
    \item[\bf Full Additivity.] For all $ \mathbf{u}, \mathbf{v}, \mathbf{w} \in \mathcal{D} $, if $\mathbf{u} \succsim \mathbf{v}$, then $\mathbf{u} +  \mathbf{w} \succsim  \mathbf{v} +  \mathbf{w}$. 
\end{description}
The dominant interpretation of this axiom is about the degree of interpersonal comparability (\citealp{dAspremont1977equity,roberts1980interpersonal,BBD2002BOOK}): This axiom has been interpreted as requiring that even if each individual's origin is changed by $w_i$, the social ranking should not be affected, i.e., prohibiting the interpersonal comparison of utility levels. 

Another interpretation is also possible (e.g., \citealp{weymark1981generalized}). It can be deemed as  postulating the invariance property of rankings under the addition of an
arbitrary new utility stream: More precisely, \textit{full additivity} can be interpreted as requiring that if $\mathbf{u}$ is weakly better than  $\mathbf{v}$, then since $\mathbf{w}$ is also weakly better than  $\mathbf{w}$ itself, the combination $\mathbf{u} + \mathbf{w}$ of weakly better ones should be at least as good as the combination $\mathbf{v} + \mathbf{w}$ of weakly worse ones. 
This interpretation is valid under fully interpersonal comparability.

We also consider weaker conditions, which can be naturally interpreted, especially under the second interpretation.
The first one requires the above independence property only when one generation's utility levels change. 
\begin{description}
    \item[\bf One-Generation Additivity.] For all $ \mathbf{u}, \mathbf{v} \in \mathcal{D} $, all $t\in \mathbb{N}$ and all $\alpha \in \mathbb{R}$, if $\mathbf{u} \succsim \mathbf{v}$, then $\mathbf{u} + \alpha  \mathbf{1}_{\{ t \}} \succsim  \mathbf{v} +  \alpha  \mathbf{1}_{\{ t \}}$. 
\end{description}

In the finite-population case, the corresponding axiom was introduced in \citet{mariotti2013impossibility}. In this case, applying the additivity axiom repeatedly yields \textit{full additivity}; however, since this paper studies the infinite-population case, \textit{one-generation additivity} is distinct from \textit{full additivity}. 

The next one considers only periodic streams.
By focusing on these streams, we can limit the number of utility levels involved.  
\begin{description}
    \item[\bf Periodic Additivity.] For all periodic sequences $\mathbf{u}, \mathbf{v}, \mathbf{w} \in \mathcal{D} $, if $\mathbf{u} \succsim \mathbf{v}$, then $\mathbf{u} +  \mathbf{w} \succsim  \mathbf{v} +  \mathbf{w}$. 
\end{description}

Finally, we introduce axioms that require that the evaluations of utility streams can be approximated from the limiting behavior of the evaluations of their $T$-heads as $T$ goes to infinity. 
Consider two utility streams $\mathbf{u}$ and  $\mathbf{v}$. 
These axioms postulate that if, for all sufficiently large $T$, the utility vector consisting of the first $T$ generations in $\mathbf{u}$ is weakly better than the corresponding vector in $\mathbf{v}$, then the overall value of $\mathbf{u}$ is weakly higher than that of $\mathbf{v}$. 
When formalizing such requirements, an important issue concerns how the evaluation of a part of a utility stream is derived from the original stream.

To address this, many consistency axioms studied in the literature replace the tails of utility streams with common utility streams (e.g., \citealp{FM2003JME,AT2004ET,asheim2010generalized,S2010SCW}). 
For instance, \citeauthor{FM2003JME}'s (\citeyear{FM2003JME}) consistency axiom requires that for all $ \mathbf{u},  \mathbf{v} \in \mathcal{D}$, if there exist $\mathbf{w} \in \mathcal{D}$ and $T^\ast  \in \mathbb{N}$ such that $( \mathbf{u}_{[1, \cdots,T ]}, \mathbf{w}_{[T+1, \cdots,\infty )})   \succsim ( \mathbf{v}_{[1, \cdots,T ]}, \mathbf{w}_{[T+1, \cdots,\infty )}) $ for all $T \geq T^\ast$, then $\mathbf{u} \succsim  \mathbf{v}$.
Although such formalizations are widely adopted, they are not well-suited to capturing key properties of social welfare orderings that primarily focus on the limiting behavior of utility streams.
To see this, consider a social welfare ordering $\succsim$ such that for all  $\mathbf{u}, \mathbf{v} \in \mathcal{D}$, if there exists $T\in\mathbb{N}$ such that $\mathbf{u}_{[T+1, \cdots,\infty )} = \mathbf{v}_{[T+1, \cdots,\infty )}$, then $ \mathbf{u} \sim \mathbf{v}$.\footnote{
For instance, social welfare orderings that are represented by a tail-monotone continuous function satisfy this property.
}
The premise of \citeauthor{FM2003JME}'s (\citeyear{FM2003JME}) axiom holds for every pair $(\mathbf{u},  \mathbf{v} ) \in \mathcal{D}^2$, which implies that all utility streams are indifferent under this ordering.
This observation suggests that to capture the implications of such orderings, the tails of utility streams should instead be replaced by values that reflect information from the first finitely many generations.

In the first axiom, the characteristics of the initial segment are incorporated into the limiting behavior by replacing the tail with the average utility level of the first finite generations.
That is, it requires that if the planner evaluates $(\mathbf{u}_{[1, \cdots,T ]}, \mu_T ( \mathbf{u}) \mathbf{1}_\mathbb{N})$ to be weakly better than $(\mathbf{v}_{[1, \cdots,T ]}, \mu_T ( \mathbf{v}) \mathbf{1}_\mathbb{N})$ for all sufficiently large $T$, then $\mathbf{u}$ should also be at least as desirable as $\mathbf{v}$. 
Although the use of averages is not free from criticism, it is one of the natural ways to proxy the missing values.

\begin{description}
    \item[\bf Mean Consistency.] For all $ \mathbf{u},  \mathbf{v} \in \mathcal{D}$, if there exists $T^\ast  \in \mathbb{N}$ such that $(\mathbf{u}_{[1, \cdots,T ]}, \mu_T ( \mathbf{u}) \mathbf{1}_\mathbb{N}) \succsim ( \mathbf{v}_{[1, \cdots,T ]}, \mu_T ( \mathbf{v}) \mathbf{1}_\mathbb{N}) $ for all $T \geq T^\ast$, then $\mathbf{u} \succsim  \mathbf{v}$.
\end{description}

It should be noted that \textit{mean consistency} is compatible with not only the rules based on the Cesàro average but also many social welfare (quasi-)orderings. 
Social welfare orderings that evaluate utility streams by their infimum or supremum satisfy \textit{mean consistency}. Furthermore, any social welfare ordering that can be represented as a convex combination of the infimum, the supremum, and the Cesàro average of utility streams is compatible with this axiom.  
Thus, this consistency axiom does not directly yield the Cesàro average functions.

In the second consistency axiom, the planner considers periodic streams consisting of generations from 1 to $T$ to evaluate the $T$-head of utility streams (i.e., we implicitly assume that the comparison between $\mathbf{u}_{[1, \cdots,T ]}$ and $\mathbf{v}_{[1, \cdots,T ]}$ is represented by that between $[\mathbf{u}_{[1, \cdots,T ]}]_\text{rep}$ and $[\mathbf{v}_{[1, \cdots,T ]}]_\text{rep}$). 
Roughly speaking, the following axiom states that the evaluation of any utility stream $\mathbf{u}$ can be approximated by the asymptotic evaluation of $[\mathbf{u}_{[1, \cdots,T ]}]_\text{rep}$ as $T$ goes to infinity.

\begin{description}
    \item[\bf Replication Consistency.] For all $ \mathbf{u},  \mathbf{v} \in \mathcal{D}$, if there exists $T^\ast  \in \mathbb{N}$ such that $[\mathbf{u}_{[1, \cdots,T ]}]_\text{rep} \succsim [\mathbf{v}_{[1, \cdots,T ]}]_\text{rep}$ for all $T \geq T^\ast$, then $\mathbf{u} \succsim  \mathbf{v}$. 
\end{description}

We consider a stronger version. The following requires the consistency property if $[\mathbf{u}_{[1, \cdots,T ]}]_\text{rep} $ is weakly better than $[\mathbf{v}_{[1, \cdots,T ]}]_\text{rep}$ periodically. 

\begin{description}
    \item[\bf Fixed-Step Replication Consistency.] For all $ \mathbf{u},  \mathbf{v} \in \mathcal{D}$, if there exists $k  \in \mathbb{N}$ such that $[\mathbf{u}_{[1, \cdots,kT ]}]_\text{rep} \succsim [\mathbf{v}_{[1, \cdots,kT ]}]_\text{rep}$ for all $T \in\mathbb{N}$, then $\mathbf{u} \succsim  \mathbf{v}$. 
\end{description}

Similar fixed-step extensions of consistency axioms have been considered by \citet{FM2003JME}, \citet{KK2009SCW}, and \citet{AB2010IJET}. 
We use the fixed-step version since it provides simple characterizations in  the larger domain $\ell^\infty$. 
The first result on $\ell^\text{Ces}$ is invariant if we impose \textit{replication consistency} instead of \textit{fixed-step replication consistency}. 
We will discuss  later how other results change when we replace \textit{fixed-step replication consistency} with \textit{replication consistency} (cf.~Remark \ref{rem_axiom}).

\section{Characterization Results}
\label{sec_main}
\subsection{The Cesàro Average Function on the Restricted Domain}
\label{sec_chara_rest}

First, we characterize the social welfare ordering that evaluates all utility streams by their Cesàro average on the restricted domain  $\ell^\text{Ces}$.

\begin{theorem}
\label{thm_chara_lCes}
    Let $\succsim$ be a social welfare ordering on $\ell^\text{Ces}$. Then the following statements are equivalent: 
    \begin{enumerate}
        \item It satisfies \textit{uniform Pareto}, \textit{finite anonymity}, \textit{continuity}, \textit{one-generation additivity}, and \textit{mean consistency}.
        \item It satisfies \textit{uniform Pareto}, \textit{periodic anonymity}, \textit{continuity}, \textit{periodic additivity}, and \textit{fixed-step replication consistency}.
        \item It is the Cesàro average social welfare ordering. 
    \end{enumerate}
\end{theorem}

Compared to \citet{P2022JET} and \citet{li2024simple}, we provide an alternative characterization of the Cesàro average social welfare ordering by directly considering utility streams instead of streams of objects. 
While \citet{P2022JET} and \citet{li2024simple} used variants of the separability axiom to characterize this rule, we obtain the characterizations by extending the additivity condition to the infinite-dimensional setup.\footnote{The axiom of separability requires that when comparing two utility vectors, the utility levels of the individuals who attain the same utility level in the two vectors should not influence the comparison.} 
Furthermore, the new consistency axioms play key roles in our characterization. 
For instance, \textit{mean consistency} excludes many rules, such as the long-run egalitarian ordering represented by the function $W: \ell^\text{Ces} \to \mathbb{R}$ such that $W(\mathbf{u}) = \lim \inf_{t\to \infty } u_t$. 
See also Appendix \ref{app_ind} for \textit{fixed-step replication consistency}. 

Before providing a proof of this theorem, we examine the implications of the axioms of anonymity and additivity in the statements (1)  and (2), respectively. 

\begin{lem}
\label{lem_ADDIE}
Suppose that  a social welfare ordering $\succsim$  on $\mathcal{D}$ satisfies \textit{finite anonymity} and \textit{one-generation additivity}. 
Then for all $\mathbf{u}, \mathbf{v} \in \mathcal{D}$, if there exist $s, t \in \mathbb{N}$ such that $u_s + u_t = v_s + v_t$ and $u_i = v_i$ for all $i \in \mathbb{N}\backslash \{s, t\}$, then $ \mathbf{u} \sim  \mathbf{v}$. 
\end{lem}

\begin{proof}
    Suppose that there exist $s, t \in \mathbb{N}$ such that $u_s + u_t = v_s + v_t$ and $u_i = v_i$ for all $i \in \mathbb{N}\backslash \{s, t\}$. 
    Define $\beta$ as  
    \begin{equation*}
        \beta ={u_s + v_s \over 2} - {u_t + v_t  \over 2}. 
    \end{equation*}
    Consider the two utility streams $\mathbf{u} + \beta\mathbf{1}_{\{ t \}}$ and $\mathbf{v} + \beta  \mathbf{1}_{\{ t \}}$. 
    By the definition of $\beta$, we have 
    \begin{align*}
        u_t + \beta &= u_t + {u_s + v_s \over 2} - {u_t + v_t  \over 2} = v_s + {u_s + u_t - (v_s + v_t) \over 2} = v_s, 
    \end{align*}
    where the last equality follows from $u_s + u_t = v_s + v_t$. 
    In the same way, we obtain $u_s = v_t + \beta$. 
    Since $u_i = v_i$ for all $i \in \mathbb{N}\backslash \{s, t\}$,  
    by \textit{finite anonymity}, we have $\mathbf{u} + \beta\mathbf{1}_{\{ t \}} \sim  \mathbf{v} + \beta\mathbf{1}_{\{ t \}}$. 
    By \textit{one-generation additivity}, we obtain $\mathbf{u} \sim \mathbf{v}$, as required. 
\end{proof}

Similarly, we can prove the following lemma. 

\begin{lem}
\label{lem_ADDIE2}
Suppose that  a social welfare ordering $\succsim$  on $\mathcal{D}$ satisfies \textit{periodic anonymity} and \textit{periodic additivity}. 
Then for all $T\in\mathbb{N}$ and all $[\mathbf{u}_{[1, \cdots, T]}]_\text{rep}, [\mathbf{v}_{[1, \cdots, T]}]_\text{rep} \in \mathcal{D}$, if there exist $s, t \in \{1, \cdots, T \}$ such that  $u_s + u_t = v_s + v_t$ and $u_i = v_i$ for all $i \in \{ 1,\cdots, T \} \backslash \{s, t\}$, then $ [\mathbf{u}_{[1, \cdots, T]}]_\text{rep} \sim [\mathbf{v}_{[1, \cdots, T]}]_\text{rep}$. 
\end{lem}

Then we provide a proof of Theorem \ref{thm_chara_lCes}.

\begin{proof}[Proof of Theorem \ref{thm_chara_lCes}]
    ``(1)$\implies$(3).'' 
    Let $\succsim$ be a social welfare ordering that satisfies the five axioms in (1). 
    First, we prove that $\mu_\infty (\mathbf{u}) > \mu_\infty (\mathbf{v})$ implies $ \mathbf{u} \succ \mathbf{v}$. 
    Let $\varepsilon \in (0,  \mu_\infty (\mathbf{u}) - \mu_\infty (\mathbf{v}))$ and define $ \mathbf{w}\in \ell^\text{Ces}$ by $w_t = u_t - \varepsilon$ for all $t \in \mathbb{N}$. 
    Note  that by \textit{uniform Pareto}, $ \mathbf{u} \succ  \mathbf{w}$. 
    Since $\mu_\infty (\mathbf{w}) > \mu_\infty ( \mathbf{v})$, there exists $T^\ast \in \mathbb{N}$ such that for all $T \geq T^\ast$, $\mu_T ( \mathbf{w}) > \mu_T (\mathbf{v})$. 
    For each $T \geq T^\ast$, consider the two utility streams $(\mathbf{v}_{[1, \cdots, T]}, \mu_T (\mathbf{v}) \mathbf{1}_\mathbb{N} )$ and $(\mathbf{w}_{[1, \cdots, T]}, \mu_T (\mathbf{w}) \mathbf{1}_\mathbb{N} )$.
    Let $\delta = \mu_T (\mathbf{w}) -  \mu_T  ( \mathbf{v})$. 
    Let $\mathbf{z}^{0} = (\mathbf{v}_{[1, \cdots, T]}, \mu_T (\mathbf{v}) \mathbf{1}_\mathbb{N} )$ and for all $t \in \{ 1,2, \cdots , T-1 \}$, 
    define $ \mathbf{z}^t$ by 
    \begin{align*}
        &z^t_s = z^{t-1}_s ~~~~ \text{for all $s \in \mathbb{N} \backslash \{ t , t+1 \}$}, \\
        &z^t_{t} = w_{t} - \delta,  \\
        &z^t_{t+1} = z^{t-1}_{t+1} - w_{t} + \delta + z^{t-1}_{t} .
    \end{align*}
    By Lemma \ref{lem_ADDIE},  $\mathbf{z}^{t-1} \sim  \mathbf{z}^{t}$ for all $t  \in \{ 1, \cdots, T-1\}$. 
    By transitivity, we have $ \mathbf{z}^{T-1} \sim  \mathbf{z}^{0} = (\mathbf{v}_{[1, \cdots, T]}, \mu_T (\mathbf{v}) \mathbf{1}_\mathbb{N} )$. 
    Note that by construction, $z^{T-1}_i = w_{i} - \delta $ for all $i \in \{1 , \cdots, T \}$.
    It follows from \textit{uniform Pareto} and transitivity that $(\mathbf{w}_{[1, \cdots, T]}, \mu_T (\mathbf{w}) \mathbf{1}_\mathbb{N} ) \succ (\mathbf{v}_{[1, \cdots, T]}, \mu_T (\mathbf{v}) \mathbf{1}_\mathbb{N} )$. 
    Since this holds for each $T \geq T^\ast$, \textit{mean consistency} implies that $ \mathbf{w} \succsim \mathbf{v}$. 
    By transitivity, we obtain $ \mathbf{u} \succ  \mathbf{v}$. 

     Next, we prove that if $\mu_\infty ( \mathbf{u}) \geq \mu_\infty ( \mathbf{v})$, then $\mathbf{u} \succsim \mathbf{v}$. For all $k \in \mathbb{N}$, define $ \mathbf{w}^k \in \ell^\text{Ces}$ by $\mathbf{w}^k = \mathbf{u} + (1/k) \mathbf{1}_\mathbb{N}$. 
    Since $\mu_\infty (\mathbf{w}) = \mu_\infty ( \mathbf{u}) + 1/k > \mu_\infty ( \mathbf{v})$, the result of  the last paragraph implies that $ \mathbf{w}^k \succ \mathbf{v}$ for all $k \in \mathbb{N}$. 
    Note that the sequence $\{ \mathbf{w}^k \}_{k \in \mathbb{N}}$ converges to $\mathbf{u}$. 
    By \textit{continuity}, we obtain $\mathbf{u} \succsim \mathbf{v}$.

    ``(2)$\implies$(3).'' This part can be proven in a similar way. 
     Let $\succsim$ be a social welfare ordering that satisfies the five axioms in (2). 
     First, we prove that $\mu_\infty (\mathbf{u}) > \mu_\infty (\mathbf{v})$ implies $ \mathbf{u} \succ \mathbf{v}$. 
     Let $\varepsilon \in (0, \mu_\infty (\mathbf{u}) - \mu_\infty (\mathbf{v}))$ and define $ \mathbf{w}\in \ell^\text{Ces}$ by $w_t = u_t - \varepsilon$ for all $t \in \mathbb{N}$. 
    Note  that by \textit{uniform Pareto}, $ \mathbf{u} \succ  \mathbf{w}$. 
     Since $\mu_\infty (\mathbf{w}) > \mu_\infty ( \mathbf{v})$, there exists $T^\ast \in \mathbb{N}$ such that for all $T \geq T^\ast$, $\mu_T ( \mathbf{w}) > \mu_T (\mathbf{v})$.
     For each $T \geq T^\ast$, consider the two utility streams $[\mathbf{v}_{[1, \cdots, T]}]_\text{rep}$ and $[\mathbf{w}_{[1, \cdots, T]}]_\text{rep}$.
    Let $\delta = \mu_T (\mathbf{w}) -  \mu_T  ( \mathbf{v})$. 
    Let $\mathbf{z}^{0} = [\mathbf{v}_{[1, \cdots, T]}]_\text{rep}$ and for all $t \in \{ 1,2, \cdots , T-1 \}$, 
    define $ \mathbf{z}^t$  as for all $s\in\mathbb{N}$, 
    \begin{align*}
        z^t_{s} = 
        \left\{
        \begin{array}{ll}
         w_{s} - \delta & \text{if $s\in \{ t,  T+ t, 2T+ t, \cdots \}$}, \\
        z^{t-1}_{s} - w_{s-1} + \delta + z^{t-1}_{s-1} & \text{if $s\in \{ t+ 1,  T+ t+ 1, 2T+ t + 1, \cdots \}$}, \\
        z^{t-1}_s & \text{otherwise.}
        \end{array}
        \right.
    \end{align*} 
    By Lemma \ref{lem_ADDIE2},  $\mathbf{z}^{t-1} \sim  \mathbf{z}^{t}$ for all $t  \in \{ 1, \cdots, T-1\}$. 
    By transitivity, we have $ \mathbf{z}^{T-1} \sim  \mathbf{z}^{0} = [\mathbf{v}_{[1, \cdots, T]}]_\text{rep}$. 
    Note that by construction, $z^{T-1}_i = w_{i} - \delta $ for all $i \in \mathbb{N}$.
    It follows from \textit{uniform Pareto} and transitivity that $[\mathbf{w}_{[1, \cdots, T]}]_\text{rep} \succ [\mathbf{v}_{[1, \cdots, T]}]_\text{rep}$. 
    Since this holds for each $T \geq T^\ast$, \textit{fixed-step replication consistency} implies that $ \mathbf{w} \succsim \mathbf{v}$. 
    By transitivity, we obtain $ \mathbf{u} \succ  \mathbf{v}$. 

    Similarly, we can prove that  if $\mu_\infty ( \mathbf{u}) \geq \mu_\infty ( \mathbf{v})$, then $\mathbf{u} \succsim \mathbf{v}$ using \textit{continuity}.

    ``(3)$\implies$(1).''
    We only prove that the Cesàro average social welfare ordering on $\ell^\text{Ces}$ satisfies \textit{mean consistency}. 
    Suppose to the contrary that there exists $T^\ast  \in \mathbb{N}$ such that $(\mathbf{u}_{[1, \cdots,T ]}, \mu_T ( \mathbf{u}) \mathbf{1}_\mathbb{N}) \succsim ( \mathbf{v}_{[1, \cdots,T ]}, \mu_T ( \mathbf{v}) \mathbf{1}_\mathbb{N}) $ for all $T \geq T^\ast$ but $\mathbf{v} \succ \mathbf{u}$.
    Since $\succsim$ is the Cesàro average social welfare ordering, $\mu_\infty (\mathbf{v}) > \mu_\infty (\mathbf{u})$. 
    Therefore, there exists $t^\ast \in\mathbb{N}$ such that for all $t \geq t^\ast$, $\mu_{t} (\mathbf{v}) > \mu_{t} (\mathbf{u})$, which implies $(\mathbf{v}_{[1, \cdots, t]}, \mu_t ( \mathbf{v}) \mathbf{1}_\mathbb{N}) \succ ( \mathbf{u}_{[1, \cdots,t ]}, \mu_t ( \mathbf{u}) \mathbf{1}_\mathbb{N})$. 
    This is a contradiction.

    ``(3)$\implies$(2).''
    We only prove that the Cesàro average social welfare ordering on $\ell^\text{Ces}$ satisfies \textit{fixed-step replication consistency}.
    Suppose to the contrary that there exists $k\in\mathbb{N}$ such that for all $T\in \mathbb{N}$, $ [\mathbf{u}_{[1, \cdots, kT]}]_\text{rep} \succsim [\mathbf{v}_{[1, \cdots, kT]}]_\text{rep}$ but $\mathbf{v} \succ \mathbf{u}$. 
    Since $\succsim$ is the Cesàro average social welfare ordering, $\mu_\infty (\mathbf{v}) > \mu_\infty (\mathbf{u})$. 
    Therefore, there exists $t^\ast \in\mathbb{N}$ such that for all $t \geq t^\ast$, $\mu_{k t} (\mathbf{v}) > \mu_{k t} (\mathbf{u})$, which implies $[\mathbf{v}_{[1, \cdots, k t]}]_\text{rep} \succ [\mathbf{u}_{[1, \cdots, k t]}]_\text{rep}$. 
    This is a contradiction. 
\end{proof}

Note that \textit{finite anonymity} and \textit{one-generation additivity} are only used in the proof of  Lemma \ref{lem_ADDIE}. 
This suggests that we can characterize the Cesàro average social welfare ordering by directly imposing their implication. 
This property was first introduced by \citet{BBD2002BOOK} as a requirement for social welfare evaluations on finite-dimensional utility vectors. 

\begin{description}
    \item[\bf Incremental Equity] For all $\mathbf{u} \in \mathcal{D}$, $i,j\in \mathbb{N}$ and $\varepsilon > 0$, $\mathbf{u} + \varepsilon \mathbf{1}_{\{ i \}} \sim \mathbf{u} + \varepsilon \mathbf{1}_{\{ j\}} $.\footnote{In welfare criteria over infinite  utility streams, \citet{KK2009SCW} considered the same axiom to characterize an extension of the utilitarian rule.} 
\end{description}

This axiom requires that when we give utility $\varepsilon$ to a generation, who obtains this utility does not matter.
In the same way as the proof of Theorem \ref{thm_chara_lCes}, the following result can be established. 

\begin{prop}
\label{cor_resCes2}
    A social welfare ordering $\succsim$ on $\ell^\text{Ces}$ satisfies \textit{uniform Pareto}, \textit{continuity},  \textit{mean consistency}, and \textit{incremental equity} if and only if it is the Cesàro average social welfare ordering. 
\end{prop}

\subsection{Direct Extensions to the Large Domain}
\label{sec_genCes}

In the previous section, we characterized the Cesàro average social welfare ordering in the restricted domain $\ell^\text{Ces}$. 
This section considers the more general domain $\ell^\infty$ and examines what class of criteria is characterized by these axioms. We verify the independence of the axioms in Appendix \ref{app_ind}.

Before stating our results, we introduce two social welfare quasi-orderings. 
The catching-up criterion $\succsim^C $ is a social welfare quasi-ordering such that for all $ \mathbf{u},  \mathbf{v} \in \mathcal{D}$, 
\begin{equation*}
     \mathbf{u} \succsim^C  \mathbf{v} 
    \iff 
    \exists T^\ast \in \mathbb{N} ~~ \text{s.t.} ~~ \forall T \geq T^\ast, ~~ \sum_{t = 1}^{T} u_t \geq \sum_{t = 1}^{T} v_t
\end{equation*}
(\citealp{atsumi1965neoclassical,von1965existence}).
This means that if for all sufficiently large $T$, the total utility level of $\mathbf{u}_{[1,\cdots, T]}$ is at least as large as that of $\mathbf{v}_{[1,\cdots, T]}$, then $\mathbf{u}$ is weakly better than $\mathbf{v}$. That is, it compares utility streams by the long-run sums. 
The \textit{fixed-step catching-up criterion} $\succsim^{\text{fix}-C}$ is a social welfare quasi-ordering such that for all $ \mathbf{u},  \mathbf{v} \in \mathcal{D}$, 
\begin{equation*}
     \mathbf{u} \succsim^{\text{fix}-C} \mathbf{v} 
    \iff 
    \exists k \in \mathbb{N} ~~ \text{s.t.} ~~ \forall T \in \mathbb{N}, ~~ \sum_{t = 1}^{kT} u_t \geq \sum_{t = 1}^{kT} v_t
\end{equation*}
(\citealp{lauwers1997infinite,FM2003JME}).
This means that if the total utility level of $\mathbf{u}_{[1,\cdots, T]}$ is at least as large as that of $\mathbf{v}_{[1,\cdots, T]}$ periodically, then $\mathbf{u}$ is weakly more desirable than $\mathbf{v}$.
Note that for all $\mathbf{u}, \mathbf{v}\in\mathcal{D}$, 
\begin{equation}
\label{eq_C-FixC}
    \mathbf{u} \succsim^C  \mathbf{v} \implies \mathbf{u} \succsim^{\text{fix}-C} \mathbf{v} 
\end{equation}
and that the converse does not hold. For instance, consider the sequences $\mathbf{u} = (1,0,1,0,\cdots)$ and $\mathbf{v} = (0,1,0,1,\cdots)$. It is easy to verify that $\mathbf{u} \succ^C \mathbf{v}$ but $\mathbf{u} \sim^{\text{fix}-C} \mathbf{v}$.

The following result clarifies what class of social welfare orderings on the more general domain $\ell^\infty$ satisfies the axioms in Theorem \ref{thm_chara_lCes}(1). 

\begin{theorem}
\label{thm_genCes1}
     A social welfare ordering $\succsim$ on $\ell^\infty$ satisfies \textit{finite anonymity}, \textit{uniform Pareto}, \textit{continuity}, \textit{one-generation additivity}, and \textit{mean consistency} if and only if it is represented by a continuous, tail-monotone Cesàro average function respecting $\succsim^C$. 
\end{theorem}

This theorem provides guidance for evaluating utility streams outside $\ell^\text{Ces}$. 
It states that if generation $T$ considers $\mathbf{u}$ to be sufficiently better than $\mathbf{v}$ for all $T$ large enough, then $\mathbf{u}$ should be strictly better than $\mathbf{v}$, that is, the rankings should respect long-run dominance. 
Furthermore, they should be compatible with comparisons by the long-run sums. 

Then we prove Theorem \ref{thm_genCes1}. 
\citet{S2016SCW}  characterized  the social welfare orderings that satisfy \textit{uniform Pareto}, \textit{finite anonymity}, \textit{continuity}, and the following axiom.
\begin{description}
    \item[\bf Weak Non-Substitution.] For all $ \mathbf{u}, \mathbf{v} \in \mathcal{D}$, if $u_1 < v_1$ and there exists $\varepsilon > 0$ such that $u_t = v_t + \varepsilon $ for all $t\geq 2$, then $ \mathbf{u} \succsim \mathbf{v}$. 
\end{description}
\begin{lem}[Theorem 1 of \citealp{S2016SCW}]
\label{thm_Sakai}
    A social welfare ordering $\succsim$ on $\ell^\infty$ satisfies \textit{uniform Pareto}, \textit{finite anonymity}, \textit{continuity} and \textit{weak non-substitution} if and only if it is represented by a continuous, tail-monotone function $W$ such that for all $\mathbf{u} \in \ell^\infty$ if there exists $\lim_{t \rightarrow \infty} u_t$, then $W( \mathbf{u}) = \lim_{t \rightarrow \infty} u_t$.\footnote{\label{foot_head}\citet{S2016SCW} derived an additional property called "head insensitivity." This requires that for all $ \mathbf{u},  \mathbf{v} \in \ell^\infty$, there exists $s \in \mathbb{N}$  such that $u_t = v_t $ for all $t \geq s$, then $W( \mathbf{u}) =  W( \mathbf{v})$. 
    However, this is redundant. Indeed, for all $k \in \mathbb{N}$, the tail-monotonicity implies $W( \mathbf{u} + {1 \over k}\mathbf{1}_\mathbb{N}) > W( \mathbf{v})$. Since $\mathbf{u} + {1 \over k}\mathbf{1}_\mathbb{N}$ converges to $ \mathbf{u}$ as $k$ goes to $\infty$, the continuity of $W$ implies  $W( \mathbf{u}) \geq W( \mathbf{v})$. In the same way, we can prove $W( \mathbf{u}) \leq W( \mathbf{v})$. 
    Thus, we obtain $W( \mathbf{u}) = W( \mathbf{v})$, as required. 
    }
\end{lem}

\begin{proof}[Proof of Theorem \ref{thm_genCes1}]
    `If.' Let $\succsim$ be a social welfare ordering that is  represented by a continuous, tail-monotone generalized Cesàro  social welfare function $W$ respecting $\succsim^C$. 
    It is easy to prove that $\succsim$ satisfies  \textit{uniform Pareto}, \textit{one-generation additivity} and \textit{continuity} using the tail-monotonicity and continuity of $W$. 
    (Note that the tail-monotonicity and continuity implies head insensitivity,  which implies \textit{one-generation additivity}.
    For the definition of head insensitivity, see Footnote \ref{foot_head}.)

    We prove that $\succsim$ satisfies \textit{mean consistency}. 
    Suppose that there exists $T^\ast \in \mathbb{N}$ such that for all $T \geq T^\ast$, $(  \mathbf{u}_{[1,\cdots, T ]}, \mu_T ( \mathbf{u}) \mathbf{1}_\mathbb{N} ) \succsim (  \mathbf{v}_{[1,\cdots, T ]}, \mu_T ( \mathbf{v})\mathbf{1}_\mathbb{N}  )$.
    By limit selection of $W$, $\mu_T (\mathbf{u}) =  W(   \mathbf{u}_{[1,\cdots, T ]}, \mu_T ( \mathbf{u})\mathbf{1}_\mathbb{N}  ) \geq W (  \mathbf{v}_{[1,\cdots, T ]}, \mu_T ( \mathbf{v})\mathbf{1}_\mathbb{N}   ) = \mu_T ( \mathbf{v})$ holds for all $T \geq T^\ast$, which is equivalent to $\sum_{t = 1 }^{T} u_t \geq \sum_{t = 1 }^{T} v_t$ for all $T \geq T^\ast$.
    By the definition of $\succsim^C$, we have $ \mathbf{u} \succsim^C  \mathbf{v}$.
    Since $W$ respects $\succsim^C$, we obtain $\mathbf{u} \succsim  \mathbf{v}$.
    Therefore, $\succsim$ satisfies \textit{mean consistency}.  

     `Only if.'
    Let $\succsim$ be a social welfare ordering on $\ell^\infty$ that satisfies \textit{uniform Pareto}, \textit{finite anonymity}, \textit{continuity}, \textit{one-generation additivity}, and \textit{mean consistency}. 

    First, we prove that $\succsim$ satisfies \textit{weak non-substitution}. 
    Let $\mathbf{u}, \mathbf{v}\in \ell^\infty$ be such that  $u_1 < v_1$ and for some $\varepsilon > 0$,  $\mathbf{u}_{[2, \cdots, \infty)} = \mathbf{v}_{[2, \cdots, \infty)} + \varepsilon \mathbf{1}_\mathbb{N}$. 
    Let $\delta= v_1 - u_1$ and 
     $m\in \mathbb{N}$ be such that $(m-1)\varepsilon/2 \leq \delta < m\varepsilon/2$. 
    Set $\mathbf{z}^1 = \mathbf{u}$ and for all $k \in \{ 2,3,  \cdots, m+1, m+ 2 \}$, define $\mathbf{z}^k$ by 
    \begin{align*}
        &z^k_1 = z^{k-1}_1 + \varepsilon/2,  \\
        &z^k_k = z^{k-1}_k - \varepsilon/2,  \\
        &z^k_i = z^{k-1}_i  ~~~~ \text{for all $i \in \mathbb{N}\backslash \{ 1, k\}$}. 
    \end{align*}
    By Lemma \ref{lem_ADDIE}, $\mathbf{z}^{k} \sim \mathbf{z}^{k+ 1}$ for all $k\in \{1 ,2 , \cdots, m+2 \}$. 
    By transitivity, we obtain  $\mathbf{u} =  \mathbf{z}^{1} \sim \mathbf{z}^{m + 2}$. 
    Note that $z^{m+2}_1 = u_1 + m \varepsilon/2  + \varepsilon/2 > u_1 + \delta  + \varepsilon/2 = v_1 +  \varepsilon/2$ and $z^{m+2}_t \geq u_t - \varepsilon/2 = v_t + \varepsilon/2 $ for all $t \in \mathbb{N}\backslash \{ 1\}$. 
    Thus, \textit{uniform Pareto} implies $ \mathbf{z}^{m+2} \succ  \mathbf{v}$. 
    By transitivity, $\mathbf{u} \succ  \mathbf{v}$ holds.

    It follows from Lemma \ref{thm_Sakai} that  $\succsim$ is represented by a continuous, tail-monotone function $W$ such that for all $\mathbf{u} \in \ell^\infty$ if there exists $\lim_{t \rightarrow \infty} u_t$, then $W( \mathbf{u}) = \lim_{t \rightarrow \infty} u_t$.
    By Theorem \ref{thm_chara_lCes},  for all $ \mathbf{u}, \mathbf{v} \in \ell^\text{Ces}$,
    \begin{equation}
    \label{eq_cesiff}
        \mathbf{u} \succsim \mathbf{v}
        \iff
        \mu_\infty (\mathbf{u}) \geq \mu_\infty ( \mathbf{v}) . 
    \end{equation}
    By \eqref{eq_cesiff}, for all $ \mathbf{w} \in \ell^\text{Ces}$, 
    $W(\mathbf{w}) = W( \mu_\infty ( \mathbf{w})\mathbf{1}_\mathbb{N} ) =  \mu_\infty (\mathbf{w})$. 
    
    Finally, we prove that for all $\mathbf{u}, \mathbf{v} \in \ell^\infty$,  $\mathbf{u} \succsim^C \mathbf{v}$ implies $\mathbf{u} \succsim \mathbf{v}$ (i.e., $W(\mathbf{u})\geq W(\mathbf{v})$). 
    By the definition of $\succsim^C$, there exists $T^\ast \in \mathbb{N}$ such that for all $T \geq T^\ast$,  $\sum_{t= 1}^{T} u_t \geq \sum_{t= 1}^{T} v_t$, that is,  $\mu_T ( \mathbf{u}) \geq \mu_T ( \mathbf{v})$.
    By \eqref{eq_cesiff}, $( \mathbf{u}_{[1,\cdots, T ]}, [\mu_T ( \mathbf{u})]_\text{rep} ) \succsim (  \mathbf{v}_{[1,\cdots, T ]}, [\mu_T ( \mathbf{v})]_\text{rep})$ for all $T\geq T^\ast$. 
    By \textit{mean consistency}, we have $ \mathbf{u} \succsim \mathbf{v}$. 
\end{proof}

Next, we examine the implications of the axioms in Theorem \ref{thm_chara_lCes}(2). 

\begin{theorem}
\label{thm_genCesfixed}
     A social welfare ordering $\succsim$ on $\ell^\infty$ satisfies \textit{periodic anonymity}, \textit{uniform Pareto}, \textit{continuity}, \textit{periodic additivity}, and \textit{fixed-step replication consistency} if and only if it is represented by a continuous, weakly monotone Cesàro average  function respecting $\succsim^{\text{fix}-C}$. 
\end{theorem}

We note the differences between the class of social welfare functions characterized in the above theorem and Theorem \ref{thm_genCes1}.
Although both sets of axioms characterize the Cesàro average social welfare ordering on $\ell^\text{Ces}$ (Theorem \ref{thm_chara_lCes}), they derive different behaviors outside $\ell^\text{Ces}$. 
Since tail-monotonicity is stronger than weak monotonicity, the rankings obtained in Theorem \ref{thm_genCesfixed} are less sensitive to long-run dominance. 
In contrast, since the respect for $\succsim^{\text{fix}-C}$ implies that for $\succsim^C$, the rankings in Theorem \ref{thm_genCesfixed} are more sensitive to the long-run sums. 
Thus, there is a tradeoff between these two classes.%
\footnote{
This suggests that one of these classes would not nest the other. Establishing this relationship formally is left for future work. 
}

\begin{proof}[Proof of Theorem \ref{thm_genCesfixed}]
    `If.' Let $\succsim$ be a social welfare ordering that is represented by a continuous, weakly monotone generalized Cesàro average function $W$ respecting $\succsim^{\text{fix}-C}$. 
    It is easy to prove that $\succsim$ satisfies  \textit{periodic anonymity}, \textit{uniform Pareto}, \textit{periodic additivity}, and \textit{continuity}.

    We prove that $\succsim$ satisfies \textit{fixed-step replication consistency}. 
    Suppose that there exists $k  \in \mathbb{N}$ such that $[\mathbf{u}_{[1, \cdots,kT ]}]_\text{rep} \succsim [\mathbf{v}_{[1, \cdots,kT ]}]_\text{rep}$ for all $T \in\mathbb{N}$. 
    Since $W$ is a Cesàro average function, 
    $\mu_{kT} (\mathbf{u}) =  W( [\mathbf{u}_{[1,\cdots, kT ]}]_\text{rep} ) \geq W (  [\mathbf{v}_{[1,\cdots, kT ]}]_\text{rep} ) = \mu_{kT} ( \mathbf{v})$ holds for each $T \in\mathbb{N}$, which is equivalent to $\sum_{t = 1}^{kT} u_t \geq \sum_{t = 1 }^{kT} v_t$ for each $T \in \mathbb{N}$.
    By the definition of $\succsim^{\text{fix}-C}$, we have $ \mathbf{u} \succsim^{\text{fix}-C} \mathbf{v}$.
    Since $W$ respects $\succsim^{\text{fix}-C}$, we obtain $\mathbf{u} \succsim  \mathbf{v}$.
    Therefore, $\succsim$ satisfies \textit{fixed-step replication consistency}.  

    `Only if.'
    Let $\succsim$ be a social welfare ordering on $\ell^\infty$ that satisfies \textit{periodic anonymity}, \textit{uniform Pareto}, \textit{continuity}, \textit{periodic additivity}, and \textit{fixed-step replication consistency}. 

    By  \textit{uniform Pareto} and \textit{continuity}, for all $\mathbf{u} \in \ell^\infty$, there exists $c_\mathbf{u}$ such that $\mathbf{u} \sim c_\mathbf{u} \mathbf{1}_\mathbb{N}$. 
    Define the function $W : \ell^\infty \rightarrow \mathbb{R}$ as for all $\mathbf{u} \in \ell^\infty$, $W(\mathbf{u} ) = c_\mathbf{u}$. 
    This function is continuous and weakly monotone. 
    By the definition of $W$ and Theorem \ref{thm_chara_lCes}, 
    for all $\mathbf{u} \in \ell^\text{Ces}$, $W(\mathbf{u}) = W(\mu_\infty (\mathbf{u}) \mathbf{1}) = \mu_\infty (\mathbf{u})$, that is, $W$ is a Cesàro average function. 

    Finally, we prove that for all $\mathbf{u}, \mathbf{v} \in \ell^\infty$,  $\mathbf{u} \succsim^{\text{fix}-C} \mathbf{v}$ implies $\mathbf{u} \succsim \mathbf{v}$ (i.e., $W(\mathbf{u})\geq W(\mathbf{v})$). 
    By the definition of $\succsim^{\text{fix}-C}$, there exists $k \in \mathbb{N}$ such that for all $T  \in \mathbb{N}$,  $\sum_{t= 1}^{kT} u_t \geq \sum_{t= 1}^{kT} v_t$, that is,  $\mu_{kT} ( \mathbf{u}) \geq \mu_{kT} ( \mathbf{v})$.
    Since $\succsim$ is represented by a Cesàro average function, $ [\mathbf{u}_{[1,\cdots, kT ]}]_\text{rep} \succsim [\mathbf{v}_{[1,\cdots, kT ]}]_\text{rep} $ for all $T\in\mathbb{N}$. 
    By \textit{fixed-step replication consistency}, we have $ \mathbf{u} \succsim \mathbf{v}$. 
\end{proof}

\begin{remark}
\label{rem_exis}
    Note that there exist social welfare orderings that satisfy all the properties in Theorems \ref{thm_genCes1} and \ref{thm_genCesfixed}. 
    Appendix \ref{app_exist} shows that the functions  $W^1, W^2, W^3, W^4$ defined as for all $\mathbf{u}\in \ell^\infty$, 
    \begin{align*}
        W^1 (\mathbf{u}) & = \sup_{k\in\mathbb{N}} \liminf_{T \rightarrow \infty} \mu_{kT} ({ \mathbf{u}}), 
        \\
        W^2 (\mathbf{u}) &= \liminf_{\delta \rightarrow 1^-}  \sigma_\delta ( \mathbf{u}),
        \\
        W^3 (\mathbf{u}) &= \limsup_{\delta \rightarrow 1^-} \sigma_\delta ( \mathbf{u}), 
        \\
        W^4 (\mathbf{u}) &= \inf_{k\in\mathbb{N}} \limsup_{T \rightarrow \infty} \mu_{kT} (\mathbf{u})
    \end{align*}
    satisfy  all of the conditions. 
\end{remark}

\subsection{Fully Additive Extensions}
\label{sec_fulladd}

This section examines the fully additive extensions to $\ell^\infty$ of the Cesàro average social welfare ordering defined on $\ell^\text{Ces}$. 
By imposing \textit{full additivity} in the results of the last section instead of \textit{one-generation additivity} or \textit{periodic additivity}, we obtain a linear social welfare function characterized by simple inequalities. 

First, we examine the axioms in Theorem \ref{thm_genCes1}. 

\begin{theorem}
\label{thm_full1}
    A social welfare ordering $\succsim$ satisfies \textit{uniform Pareto}, \textit{finite anonymity}, \textit{continuity}, \textit{full additivity}, and \textit{mean consistency} if and only if it is represented by a linear function $W$ such that for all $\mathbf{u}\in \ell^\infty$, 
    \begin{equation}
    \label{eq_simple1}
         \liminf_{T \rightarrow \infty}  \mu_{T} (\mathbf{u}) \leq W ( \mathbf{u}) \leq  \limsup_{T \rightarrow \infty} \mu_{T}(\mathbf{u}). 
    \end{equation}
\end{theorem}

By these inequalities, we have $W (\mathbf{u} ) = \mu_\infty (\mathbf{u})$ for all $\mathbf{u} \in \ell^\text{Ces}$, that is, $W$ is a Cesàro average function. 
This result provides upper and lower bounds, which are closely related to the Cesàro average, for the evaluations of each utility stream.

In Appendix \ref{app_exist}, we prove the existence of evaluation rules satisfying \eqref{eq_simple1} using the Hahn--Banach extension theorem, 
that is, relying on the Axiom of Choice. 
This reliance limits the applicability of these social welfare orderings because they may not be constructable. 
Examining whether the Axiom of Choice is necessary or not is left for future work. 

It should be noted that the functions derived in Theorem \ref{thm_full1} are special cases of \textit{Banach limits}, real-valued linear functions $\Lambda$ on $\ell^\infty$ satisfying the following properties: 
\begin{itemize} 
    \item For all $\mathbf{u}\in \ell^\infty$, $\Lambda( \mathbf{u}) = \Lambda( \mathbf{u}_{[2, \cdots, \infty)]} ) $. 
    \item For all $\mathbf{u}\in \ell^\infty$,  $\liminf_{t \rightarrow \infty}  u_t \leq  \Lambda( \mathbf{u}) \leq \limsup_{t \rightarrow \infty} u_t$. 
\end{itemize}
Indeed, by the linearity of $W$, we have $W(\mathbf{u}) - W(\mathbf{u}_{[2, \cdots, \infty)]}) = W(\mathbf{u} - \mathbf{u}_{[2, \cdots, \infty)]}) = \lim_{t\rightarrow \infty} {1 \over t} ( u_1 - u_{t+1} )  = 0$. The second property immediately follows from \eqref{eq_simple1}. 
Thus, Theorem \ref{thm_full1} can be interpreted as identifying an appealing subclass of Banach limits from the perspective of intergenerational fairness.

\begin{proof}[\it Proof of Theorem \ref{thm_full1}]
    Let $\succsim$ be a social welfare ordering that satisfies the five axioms. 
    By \textit{uniform Pareto} and \textit{continuity}, for all $\mathbf{u}\in\ell^\infty$, there exists a unique number $c_\mathbf{u} \in\mathbb{R}$ such that $\mathbf{u} \sim c_\mathbf{u} \mathbf{1}_\mathbb{N}$. 
    Define the function $W:\ell^\infty \rightarrow \mathbb{R}$ as for all $\mathbf{u} \in \ell^\infty$, $W(\mathbf{u}) = c_\mathbf{u}$. 
    Note that $W$ represents $\succsim$ and by \textit{uniform Pareto} and \textit{continuity}, for all $\mathbf{u}, \mathbf{v} \in\ell^\infty$,
    \begin{equation}
    \label{eq_Fmono}
        \mathbf{u} \geq \mathbf{v} \implies W (\mathbf{u}) \geq W( \mathbf{v}). 
    \end{equation}
    By Theorem \ref{thm_genCes1} and the definition of $W$, for all $\mathbf{u}\in \ell^\text{Ces}$, $W (\mathbf{u}) = W (\mu_\infty (\mathbf{u} ) \mathbf{1}_\mathbb{N}) = \mu_\infty (\mathbf{u} )$. 
    
    \vspace{3mm}
    \noindent
    Claim 1. For all $\mathbf{u}, \mathbf{v}\in \ell^\infty$, $W(\mathbf{u} + \mathbf{v}) = W(\mathbf{u} ) + W(\mathbf{v})$. 

    \begin{proof}
        By \textit{full additivity}, we have $\mathbf{u} + \mathbf{v} \sim c_\mathbf{u} \mathbf{1}_\mathbb{N} + \mathbf{v}$ and $\mathbf{v} + c_\mathbf{u} \mathbf{1}_\mathbb{N} \sim c_\mathbf{v} \mathbf{1}_\mathbb{N} + c_\mathbf{u} \mathbf{1}_\mathbb{N}$, which imply that  $W(\mathbf{u} + \mathbf{v}) = W(c_\mathbf{u} \mathbf{1}_\mathbb{N} + \mathbf{v})$ and $W (\mathbf{v} + c_\mathbf{u} \mathbf{1}_\mathbb{N})  = W(c_\mathbf{v} \mathbf{1}_\mathbb{N} + c_\mathbf{u} \mathbf{1}_\mathbb{N})$.
        Therefore, $W(\mathbf{u} + \mathbf{v}) =  W( (c_\mathbf{v} + c_\mathbf{u} )\mathbf{1}_\mathbb{N}) = c_\mathbf{v} + c_\mathbf{u} = W(\mathbf{u}) + W(\mathbf{v})$. 
     \end{proof}

    \noindent
    Claim 2. For all $\mathbf{u} \in \ell^\infty$ and $\alpha\in \mathbb{R}$, $W(\alpha \mathbf{u}) = \alpha W(\mathbf{u} )$. 

    \begin{proof}
        By Claim 1, we have $W(\mathbf{v}) = - W(- \mathbf{v})$ for all $\mathbf{v} \in \ell^\infty$. 
        Thus, it is sufficient to prove that for all $\mathbf{u} \in \ell^\infty$ with $W (\mathbf{u}) \geq 0$ and  all $\alpha \in \mathbb{R}_{++}$, $W (\alpha \mathbf{u}) = \alpha W(\mathbf{u} )$. 
        Let $\mathbf{u} \in \ell^\infty$ with $W (\mathbf{u}) \geq 0$.  
        By Claim 1, we have $W (2 \mathbf{u}) = 2 W(\mathbf{u})$. By the induction, for all $m\in\mathbb{N}$, $W (m \mathbf{u}) = mW(\mathbf{u})$. 
        
        Let $q$ be a positive rational number. Note that there exist $m,n \in \mathbb{N}$ such that $q = m/n$. 
        By the result of the last paragraph, $ W ( n q \mathbf{u} ) = W ( m\mathbf{u} )$ implies $n W ( q \mathbf{u} ) =m W ( \mathbf{u} )$, or $W ( q \mathbf{u} ) = (m/n) W ( \mathbf{u} ) = q W ( \mathbf{u} )$. 

        We prove that for all $\alpha \in \mathbb{R}_{++}$, $W (\alpha \mathbf{u}) = \alpha W (\mathbf{u} )$. Let $\{ \overline{\alpha}^k \}_{k\in\mathbb{N}}, \{ \underline{\alpha}^k \}_{k\in\mathbb{N}}$ be sequences of positive rational numbers such that $\overline{\alpha}^k \downarrow \alpha$ and $\underline{\alpha}^k \uparrow \alpha$. 
        By \eqref{eq_Fmono}, we have $ W(\overline{\alpha}^k \mathbf{u} ) \geq W(\alpha \mathbf{u}) \geq W (\underline{\alpha}^k \mathbf{u} ) $ for all $k\in\mathbb{N}$. 
        Since $\overline{\alpha}^k$ and $\underline{\alpha}^k$ are positive rational numbers, we have $\overline{\alpha}^k W ( \mathbf{u} ) \geq W (\alpha \mathbf{u}) \geq \underline{\alpha}^k W ( \mathbf{u} ) $. 
        By $\overline{\alpha}^k \downarrow \alpha$ and $\underline{\alpha}^k \uparrow \alpha$, we have $W(\alpha \mathbf{u}) = \alpha W(\mathbf{u} )$. 
     \end{proof} 

    By Claims 1 and 2, $W$ is a linear function.
    Then, we prove that $W$ satisfies the inequalities in the statement. 

    \vspace{3mm}
    \noindent
    Claim 3. For all $\mathbf{u}\in \ell^\infty$,  $\liminf_{T \rightarrow \infty}  \mu_T (\mathbf{u}) \leq W (\mathbf{u})$. 

    \begin{proof}
        Suppose to the contrary that for some $\mathbf{u}\in\ell^\infty$, $\liminf_{T \rightarrow \infty}  \mu_T (\mathbf{u}) > W ( \mathbf{u})$. 
        Let $\alpha \in (W ( \mathbf{u}), \liminf_{T \rightarrow \infty}  \mu_T (\mathbf{u}))$. 
        Then,  there exists $T^\ast$ such that for all $T\geq T^\ast$, $ \mu_T (\mathbf{u}) \geq  \mu_T (\alpha \mathbf{1}_\mathbb{N} ) $. 
        Since $W$ respects $\succsim^C$ (Theorem \ref{thm_genCes1}), $ \mathbf{u} \succsim  \alpha \mathbf{1}_\mathbb{N}$. 
        By \textit{uniform Pareto}, $\alpha \mathbf{1}_\mathbb{N} \succ W(\mathbf{u}) \mathbf{1}_\mathbb{N}$. 
        By transitivity and the definition of $W$, $\mathbf{u} \succ  W(\mathbf{u}) \mathbf{1}_\mathbb{N} \sim \mathbf{u}$, a contradiction. 
    \end{proof}

    \noindent
    Claim 4. For all $\mathbf{u}\in \ell^\infty$,  $\limsup_{T \rightarrow \infty}  \mu_T (\mathbf{u}) \geq W ( \mathbf{u})$. 

    \begin{proof} 
        Suppose to the contrary that $\limsup_{T \rightarrow \infty}  \mu_T (\mathbf{u}) < W ( \mathbf{u})$. 
        Then,  $\liminf_{T \rightarrow \infty}  \mu_T (W ( \mathbf{u}) \mathbf{1}_\mathbb{N} - \mathbf{u}) > 0$. 
        By Claim 3, $W(W ( \mathbf{u}) \mathbf{1}_\mathbb{N} - \mathbf{u}) > 0$. 
        By the definition of $W$, $W(\mathbf{u}) = W(W ( \mathbf{u}) \mathbf{1}_\mathbb{N})  > W(\mathbf{u})$, a contradiction.
    \end{proof}

    `If.' Let $\succsim$ be a social welfare ordering 
    represented by a linear function $W$ such that for all $\mathbf{u}\in \ell^\infty$, 
    \begin{equation*}
         \liminf_{T \rightarrow \infty}  \mu_{T} (\mathbf{u}) \leq W ( \mathbf{u}) \leq  \limsup_{T \rightarrow \infty} \mu_{T}(\mathbf{u}). 
    \end{equation*}
    \begin{itemize}
        \item \textit{Uniform Pareto}: Let $\mathbf{u}, \mathbf{v} \in \ell^\infty$ be such that for some $\varepsilon > 0$, $\mathbf{u} \geq  \mathbf{v } + \varepsilon \mathbf{1}_\mathbb{N}$. 
        By the inequality, $W(\mathbf{u} - \mathbf{v}  ) \geq \liminf_{T\rightarrow \infty} \mu_T (\mathbf{u} - \mathbf{v} ) > \liminf_{T\rightarrow \infty} \mu_T (\mathbf{u} - \mathbf{v}  - \varepsilon \mathbf{1}_\mathbb{N}) \geq 0$. 
        Since $W$ is linear, $W(\mathbf{u} ) > W( \mathbf{v}  )$. 

        \item  \textit{Finite Anonymity}: Let $\mathbf{u} \in \ell^\infty$  and $\pi \in \Pi^\text{fin}$. 
        Note that  there exists $T^\ast \in\mathbb{N}$ such that for all $T\geq T^\ast$, $\mu_{T} (\mathbf{u}) = \mu_{T} (\mathbf{v})$. 
        Therefore, $\lim_{T\rightarrow \infty} \mu_{T} (\mathbf{u}-\mathbf{v}) = 0$. 
        By the inequality and the linearity, we have $W(\mathbf{u} )- W(\mathbf{v})  = W(\mathbf{u}- \mathbf{v})  = 0$.

        \item  \textit{Continuity}: Note that  for all $\mathbf{u}\in \ell^\infty$, 
        \begin{equation*}
             - \sup_{T\in\mathbb{N}} | u_ T| \leq \liminf_{T \rightarrow \infty}  \mu_T (\mathbf{u}) \leq W ( \mathbf{u}) \leq \limsup_{T \rightarrow \infty} \mu_T (\mathbf{u}) \leq \sup_{T\in\mathbb{N}} | u_ T|, 
        \end{equation*}
        that is, $|W(\mathbf{u})| \leq \sup_{T\in\mathbb{N}} | u_ T|$. 
        By the linearity of $W$, for all $\mathbf{u}, \mathbf{v}\in\ell^\infty$, $|W(\mathbf{u}) - W(\mathbf{v})| = |W(\mathbf{u} - \mathbf{v})| \leq \sup_{t\in \mathbb{N}} |u_t - v_t|$. 
        Therefore, $W$ is a continuous function, which implies that $\succsim$ satisfies \textit{continuity}. 
        
        \item  \textit{Full Additivity}: It immediately follows from  the linearity of $W$. 

        \item \textit{Mean Consistency}: Let $ \mathbf{u},  \mathbf{v} \in \ell^\infty$  be such that there exists $T^\ast  \in \mathbb{N}$ such that $(\mathbf{u}_{[1, \cdots,T ]}, \mu_T ( \mathbf{u}) \mathbf{1}_\mathbb{N}) \succsim ( \mathbf{v}_{[1, \cdots,T ]}, \mu_T ( \mathbf{v}) \mathbf{1}_\mathbb{N}) $ for all $T \geq T^\ast$. 
         By \eqref{eq_simple1}, we have $\mu_T (\mathbf{u}) \geq \mu_T (\mathbf{v} ) $ for all $T \geq T^\ast$. 
        Then, we have $\liminf_{T\rightarrow \infty}\mu_{T} (\mathbf{u} - \mathbf{v}) \geq 0$. 
        By the inequality and the linearity, we have  $W(\mathbf{u}) - W(\mathbf{v} ) =  W (\mathbf{u} - \mathbf{v})\geq \liminf_{T\rightarrow \infty}\mu_T (\mathbf{u} - \mathbf{v}) \geq 0$. \qedhere
    \end{itemize}
\end{proof}

Then we consider the axioms in Theorem \ref{thm_genCesfixed}. 

\begin{theorem}
\label{thm_full2}
    A social welfare ordering $\succsim$ satisfies \textit{uniform Pareto}, \textit{periodic anonymity}, \textit{continuity}, \textit{full additivity}, and \textit{fixed-step replication consistency} if and only if it is represented by a linear function $W$ such that for all $\mathbf{u}\in \ell^\infty$, 
    \begin{equation}
    \label{eq_simple2}
        \sup_{k\in\mathbb{N}} \liminf_{T \rightarrow \infty}  \mu_{kT} (\mathbf{u}) \leq W ( \mathbf{u}) \leq \inf_{k\in\mathbb{N}} \limsup_{T \rightarrow \infty} \mu_{kT}(\mathbf{u}). 
    \end{equation}
\end{theorem}

As in Theorem \ref{thm_full1}, the inequalities in \eqref{eq_simple2} provide upper and lower bounds for the evaluations of each utility stream. The constraints in Theorem \ref{thm_full2} are related to the fixed-step catching-up criterion $\succsim^{\text{fix}-C}$  and more stringent than those in Theorem \ref{thm_full1}.
Note that the functions derived in Theorem \ref{thm_full2} are also special cases of Banach limits. 
We verify the existence of linear functions satisfying \ref{eq_simple2} using the Hahn--Banach extension theorem (and hence the Axiom of Choice) in Appendix \ref{app_exist}. 

\begin{proof}[\it Proof of Theorem \ref{thm_full2}]
    Let $\succsim$ be a social welfare ordering that satisfies the five axioms. 
    We define the function $W:\ell^\infty \rightarrow \mathbb{R}$ in the same way as the proof of Theorem \ref{thm_full1}.  
    Note that $W$ represents $\succsim$ and by \textit{uniform Pareto} and \textit{continuity}, for all $\mathbf{u}, \mathbf{v} \in\ell^\infty$,
    \begin{equation*}
    \label{eq_Fmono1}
        \mathbf{u} \geq \mathbf{v} \implies W (\mathbf{u}) \geq W( \mathbf{v}). 
    \end{equation*}
    We can also prove the linearity of $W$ in the same way as the proof of Theorem \ref{thm_full1}. 

    We claim that for all $\mathbf{u}\in \ell^\infty$,  $\sup_{k\in\mathbb{N}} \liminf_{T \rightarrow \infty}  \mu_{kT} (\mathbf{u}) \leq W ( \mathbf{u})$. 
        Suppose to the contrary that there exists $\mathbf{u}\in\ell^\infty$ such that $\sup_{k\in\mathbb{N}} \liminf_{T \rightarrow \infty}  \mu_{kT} (\mathbf{u}) > W ( \mathbf{u})$. 
        Let $\alpha \in (W ( \mathbf{u}), \sup_{k\in\mathbb{N}} \liminf_{T \rightarrow \infty}  \mu_{kT} (\mathbf{u}))$. 
        Then, there exists $k\in \mathbb{N}$ such that for all $T\in \mathbb{N}$, $ \mu_{kT} (\mathbf{u}) \geq  \mu_{kT} (\alpha \mathbf{1}_\mathbb{N} ) $. 
        Since $W$ respects $\succsim^{\text{fix}-C}$ (Theorem \ref{thm_genCesfixed}), $ \mathbf{u} \succsim  \alpha \mathbf{1}_\mathbb{N}$. 
        By \textit{uniform Pareto}, $\alpha \mathbf{1}_\mathbb{N} \succ W(\mathbf{u}) \mathbf{1}_\mathbb{N}$. 
        By transitivity and the definition of $W$, $\mathbf{u} \succ  W(\mathbf{u}) \mathbf{1}_\mathbb{N} \sim \mathbf{u}$, a contradiction.

    Then we verify that for all $\mathbf{u}\in \ell^\infty$,  $\inf_{k\in\mathbb{N}} \limsup_{T \rightarrow \infty}  \mu_{kT} (\mathbf{u}) \geq W ( \mathbf{u})$. 
        Suppose to the contrary that $\inf_{k\in\mathbb{N}}  \limsup_{T \rightarrow \infty}  \mu_{kT} (\mathbf{u}) < W ( \mathbf{u})$. 
        Then,  
        \begin{equation*}
            \sup_{k\in\mathbb{N}}  \liminf_{T \rightarrow \infty}  \mu_{kT} (W ( \mathbf{u}) \mathbf{1}_\mathbb{N} - \mathbf{u}) > 0.
        \end{equation*} 
        By the result of the last paragraph, $W(W ( \mathbf{u}) \mathbf{1}_\mathbb{N} - \mathbf{u}) > 0$. 
        By the definition of $W$ and its linearity, $W(\mathbf{u}) = W(W ( \mathbf{u}) \mathbf{1}_\mathbb{N})  > W(\mathbf{u})$, a contradiction.

    `If.' Let $\succsim$ be a social welfare ordering represented by  a linear function $W$ such that for all $\mathbf{u}\in \ell^\infty$, 
    \begin{equation*}
        \sup_{k\in\mathbb{N}} \liminf_{T \rightarrow \infty}  \mu_{kT} (\mathbf{u}) \leq W ( \mathbf{u}) \leq \inf_{k\in\mathbb{N}} \limsup_{T \rightarrow \infty} \mu_{kT}(\mathbf{u}). 
    \end{equation*} 
    
    \textit{Periodic anonymity}, \textit{Uniform Pareto} and \textit{full additivity} can be proved in the same way as the last theorem. 
    \begin{itemize}
        \item  \textit{Continuity}: Note that  for all $\mathbf{u}\in \ell^\infty$, 
        \begin{align*}
             - \sup_{T\in\mathbb{N}} | u_ T| \leq \liminf_{T \rightarrow \infty}  \mu_T &(\mathbf{u}) 
             \leq
             \sup_{k\in\mathbb{N}} \liminf_{T \rightarrow \infty}  \mu_{kT} (\mathbf{u}) \\
             &\leq 
             W ( \mathbf{u}) \leq \inf_{k\in\mathbb{N}} \limsup_{T \rightarrow \infty}  \mu_{kT} (\mathbf{u}) \leq \limsup_{T \rightarrow \infty} \mu_T (\mathbf{u}) \leq \sup_{T\in\mathbb{N}} | u_ T|, 
        \end{align*}
        that is, $|W(\mathbf{u})| \leq \sup_{T\in\mathbb{N}} | u_ T|$
        By the linearity of $W$, for all $\mathbf{u}, \mathbf{v}\in\ell^\infty$, $|W(\mathbf{u}) - W(\mathbf{v})| = |W(\mathbf{u} - \mathbf{v})| \leq \sup_{t\in \mathbb{N}} |u_t - v_t|$. 
        Therefore, $W$ is a continuous function, which implies that $\succsim$ satisfies \textit{continuity}. 
        
        \item \textit{Fixed-Step Replication Consistency}: Let $ \mathbf{u},  \mathbf{v} \in \ell^\infty$ be such that
        there exists $k  \in \mathbb{N}$ with $[\mathbf{u}_{[1, \cdots, kT ]}]_\text{rep} \succsim [\mathbf{v}_{[1, \cdots, k T ]}]_\text{rep}$ for all $T \in \mathbb{N}$.
        By \eqref{eq_simple2}, we have $\mu_{kT} (\mathbf{u}) \geq \mu_{kT} (\mathbf{v})$ for all $T \in \mathbb{N}$. 
        Then, we have $\liminf_{T\rightarrow \infty}\mu_{kT} (\mathbf{u} - \mathbf{v}) \geq 0$. 
        Therefore, we have  $W(\mathbf{u}) - W( \mathbf{v} ) =  W (\mathbf{u} - \mathbf{v})\geq \liminf_{T\rightarrow \infty}\mu_T (\mathbf{u} - \mathbf{v}) \geq 0$. \qedhere
    \end{itemize}
\end{proof}

\begin{remark}
\label{rem_axiom}
    If we replace \textit{fixed-step replication consistency} with \textit{replication consistency} in Theorem \ref{thm_full2}, then the class of orderings characterized in Theorem \ref{thm_full1} can be obtained. 
    We omit its proof because we can prove it in the same way as the proof of Theorem \ref{thm_full1}.
\end{remark}

\section{Concluding Remarks}
\label{sec_conc}

In this paper, we have axiomatically examined  Cesàro average social welfare orderings. 
We have first provided two characterizations in the restricted domain $\ell^\text{Ces}$ (Theorem \ref{thm_chara_lCes}) and then identified what class of social welfare orderings is admitted in the larger domain $\ell^\infty$.  
While the behavior of orderings outside $\ell^\text{Ces}$ depends on which axioms we impose, they both align with impartial utilitarian criteria, such as the catching-up criterion $\succsim^C$ (Theorem \ref{thm_genCes1}) and its fixed-step version $\succsim^{\text{fix}-C}$ (Theorem \ref{thm_genCesfixed}). 
Furthermore, we have examined extensions of the Cesàro average social welfare ordering on $\ell^\text{Ces}$ to $\ell^\infty$ with  full additivity. 
We have shown that these social welfare orderings can be represented by linear functions constrained by simple inequalities (Theorems \ref{thm_full1} and \ref{thm_full2}).

\appendix
\renewcommand{\thesection}{A.\arabic{section}}
\setcounter{section}{0}
\renewcommand{\theequation}{\arabic{equation}}

\section*{Appendix}

\section{Proof of the results in Section 3}

\begin{proof}[Proof of Observation \ref{obs_duCes}]
    We provide a proof in a similar way to Lemma 1 of \citet{JV2018TE}.\footnote{\citet{JV2018TE} showed the related, but different inequality as follows: For all $\mathbf{u} \in \mathcal{D}$,  
    \begin{equation*}
        \liminf_{T \rightarrow \infty} C_T ({ \mathbf{u}}) \leq \liminf_{\delta \rightarrow 1^-} \sum_{t = 1}^{\infty} \delta^{t-1} u_t \leq \limsup_{\delta \rightarrow 1^-} \sum_{t = 1}^{\infty} \delta^{t-1} u_t \leq  \limsup_{T \rightarrow \infty} C_T ({ \mathbf{u}}), 
    \end{equation*}
    where 
    \begin{equation*}
        C_t ({ \mathbf{u}}) = {\sum_{s = 1}^{t} (t-s+ 1) u_s \over t}. 
    \end{equation*}
    }
    First, we prove the first inequality. 
    Let $k\in\mathbb{N}$ and $s_t = \sum_{n = 1}^{t} u_n$ for all $t\in\mathbb{N}$. Since $u_t = s_t - s_{t-1}$ for $t \geq 2$, we have 
    \begin{align*}
    \sigma_\delta ( \mathbf{u}) &= (1 - \delta) \{u_1 + \sum_{t = 2}^{\infty} \delta^{t-1} (s_t - s_{t-1}) \} \\
    &= (1 - \delta)^2 \sum_{t = 1}^{\infty} \delta^{t-1} s_t \\
    &= (1 - \delta)^2 \sum_{t = 1}^{\infty} \delta^{t-1} t \mu_t ( \mathbf{u}). 
    \end{align*}
    Let $\lambda_k = \liminf_{T \rightarrow \infty} \mu_{kT} ( \mathbf{u})$. 
    For all $\varepsilon > 0$, there exists $T^\ast$ such that for all $T > T^\ast$, $\mu_{kT} ( \mathbf{u}) - \lambda_k >- \varepsilon$. Then we obtain
    \begin{align}
    \sigma_\delta ( \mathbf{u}) - \lambda_k 
    &= (1 - \delta)^2 \sum_{t = 1}^{\infty} \delta^{t-1} t ( \mu_t ( \mathbf{u}) - \lambda_k)\notag \\
    &\geq (1 - \delta)^2 \sum_{t = 1}^{T^\ast} \delta^{t-1} t ( \mu_t ( \mathbf{u}) - \lambda_k) - \varepsilon (1 - \delta )^2 \sum_{t= T^\ast + 1}^{\infty} \delta^{t-1} t \notag\\
     \label{eq_pflem_ldu}
     & \geq  (1 - \delta)^2 \sum_{t = 1}^{T^\ast} \delta^{t-1} t ( \mu_t ( \mathbf{u}) - \lambda_k) -  \varepsilon, 
    \end{align}
    where the last inequality follows from $\sum_{t= T^\ast+ 1}^{\infty} \delta^{t-1} t \leq \sum_{t= 1}^{\infty} \delta^{t-1} t = 1/(1-\delta)^2$. As $\delta \rightarrow 1^-$, the first term in \eqref{eq_pflem_ldu} goes to $0$. 
    Thus, we have 
    \begin{equation*}
        \liminf_{\delta \rightarrow 1^-}  \sigma_\delta ( \mathbf{u}) \geq \lambda_k = \liminf_{T \rightarrow \infty} \mu_{kT} ( \mathbf{u}). 
    \end{equation*}

    We can prove the third inequality  in the same way. The second inequality is obvious. 
\end{proof}

\begin{proof}[Proof of Observation \ref{obs_limCes}]
    Let $u_\infty = \lim_{t\rightarrow \infty} u_t$. Note that for all $\varepsilon > 0$, there exists $T^\ast \in \mathbb{N}$ such that for all $t \geq T^\ast$, $u_t - u_\infty < \varepsilon/2$. Given this $T^\ast$, there exists $T^{\ast\ast} (> T^\ast)$ such that for all $T \geq T^{\ast\ast}$, $| \sum_{t = 1}^{T^\ast} (u_t - u_\infty) /T | < \varepsilon/2$.  Thus, we have 
    \begin{align*}
        \Big| {1 \over T} \sum_{t = 1}^T u_t - u_\infty \Big| 
        &\leq \Big| {1 \over T} \sum_{t = 1}^{T^\ast} (u_t - u_\infty) \Big| + \Big| {1 \over T} \sum_{t = T^\ast + 1}^{T} (u_t - u_\infty) \Big| \\
        &\leq {\varepsilon \over 2} + {\varepsilon \over 2} \Big({T - T^\ast \over T} \Big) \\
        &< \varepsilon.  
    \end{align*}
    Therefore, we obtain $\mu_\infty (\mathbf{u}) = \lim_{t\rightarrow \infty} u_t$. 
\end{proof}

\begin{proof}[Proof of Observation \ref{obs_proLces}]
    First, we show that the set $\ell^\text{Ces}$ is convex. 
    Consider two utility streams $\mathbf{u}, \mathbf{v}\in \ell^\text{Ces}$. 
    For any $\alpha \in (0,1)$, 
    \begin{align*}
        \lim_{T\rightarrow \infty} {1 \over T} \sum_{t = 1}^{T} (\alpha u_t + (1 -\alpha) v_t) 
        &= \lim_{T\rightarrow \infty} \Bigg\{ \alpha {1 \over T} \sum_{t = 1}^{T} u_t +  (1 - \alpha) {1 \over T} \sum_{t = 1}^{T} v_t \Bigg\} \\
        &= \alpha \lim_{T\rightarrow \infty} {1 \over T} \sum_{t = 1}^{T} u_t  + (1 - \alpha) \lim_{T\rightarrow \infty} {1 \over T} \sum_{t = 1}^{T} v_t \\
        &= \alpha \mu_\infty ( \mathbf{u}) + (1 - \alpha)\mu_\infty ( \mathbf{v}). 
    \end{align*}
    Thus, $\alpha \mathbf{u}+ (1-\alpha) \mathbf{v}$ is in $\ell^\text{Ces}$. 

    Then, we show that the set $\ell^\text{Ces}$ is closed. 
    It is sufficient to prove that 
    for all sequences $\{ \mathbf{u}^k\}_{k \in \mathbb{N}} (\subset \ell^\text{Ces})$, if $\{ \mathbf{u}^k \}_{k \in \mathbb{N}}$ converges to some utility stream $ \mathbf{u} \in \ell^\infty$, then $ \mathbf{u} \in \ell^\text{Ces}$. 
    For simplicity, we write $\mu^k = \mu_\infty ( \mathbf{u}^k)$ for all $k \in \mathbb{N}$. 

    By the definition of $\ell^\infty$, there exists a positive number $a$ such that $\sup_{t \in \mathbb{N}} |u_t| \leq a$. Since $\{ \mathbf{u}^k \}_{k \in \mathbb{N}}$ converges to $ \mathbf{u}$,  there exists $K \in \mathbb{N}$ such that  $\sup_{t \in \mathbb{N}} |u^k_t| \leq a +1$ for all $k \geq K$. 
    Thus, we have $ - (a + 1)\leq \mu^k \leq  a+ 1$ for all $k\geq K$. 
    Since $\{ \mu^k \}_{k \geq K}$ is a sequence in the compact set $[- (a + 1),  a+ 1]$, there exists a convergent subsequence of $\{ \mu^k \}_{k \geq K}$. 
    We denote this subsequence by $\{ \mu^l \}_{l \in A}$, where the set $A$ is an infinite subset of $\mathbb{N}$. 
    Let $\mu$ denote the convergent point of $\{ \mu^l \}_{l \in A}$. 
    The sequence corresponding to $\{ \mu^l \}_{l \in A}$ is denoted by $\{  \mathbf{u}^l \}_{l \in A}$.

    Finally, we prove that $\mu = \lim_{T \rightarrow \infty}\mu_T ( \mathbf{u})$. 
    For all $l \in A$,
    \begin{align}
        |\mu_T ( \mathbf{u}) - \mu| 
        &= \Big| {1\over T} \sum_{t = 1}^{T} u_t - \mu \Big| \notag\\
        &\leq \label{eq_propDces}
        \Big| {1\over T} \sum_{t = 1}^{T} (u_t - u^l_t)  \Big| + \Big| {1\over T} \sum_{t = 1}^{T} u^l_t - \mu^l   \Big| + |\mu^l - \mu| . 
    \end{align}
     By construction, the sequence $\{  \mathbf{u}^l \}_{l \in A}$ converges to $ \mathbf{u}$ and $ {1\over T} \sum_{t = 1}^{T} u^l_t$ converges to $\mu^l$. 
    Hence, as $T$ goes to infinity, the first and second terms of \eqref{eq_propDces} converge to $0$. 
    We have $\lim_{T \rightarrow \infty} |\mu_T ( \mathbf{u}) - \mu| \leq |\mu^l - \mu | $. As $l$ goes to infinity, $|\mu^l - \mu|$ goes to $0$. 
    Thus, we obtain $\lim_{T \rightarrow \infty} |\mu_T ( \mathbf{u}) - \mu|  = 0$. 
\end{proof}

\section{Independence of the axioms in Theorems 1, 2, and 3}
\label{app_ind}

We verify the independence of the axioms in Theorem \ref{thm_chara_lCes}(1) and Theorem \ref{thm_genCes1}. 

\begin{itemize}
    \item  \textit{Dropping Uniform Pareto}: the social welfare ordering $\succsim$ defined as for all $\mathbf{u}, \mathbf{v} \in \mathcal{D}$, $\mathbf{u} \sim \mathbf{v}$.

    \item \textit{Dropping Finite Anonymity}: the  social welfare ordering $\succsim$ represented by the function $W: \mathcal{D} \to \mathbb{R}$ such that $W(\mathbf{u}) = u_1$ for all $\mathbf{u} \in \mathcal{D}$.

    \item \textit{Dropping Continuity}: the social welfare ordering $\succsim$ defined as for all $ \mathbf{u},  \mathbf{v} \in \mathcal{D}$, 
    \begin{align*}
        \bigg[ \exists T^\ast ~~ \text{s.t.} ~~ \forall T\geq T^\ast, ~~ \sum_{t = 1}^T u_t \geq \sum_{t = 1}^T v_t \bigg] \implies \mathbf{u} \succsim \mathbf{v}, 
        \\
        \bigg[ \exists T^\ast ~~ \text{s.t.} ~~ \forall T\geq T^\ast, ~~ \sum_{t = 1}^T u_t > \sum_{t = 1}^T v_t \bigg] \implies \mathbf{u} \succ \mathbf{v}. 
    \end{align*}
    The existence of such a social welfare ordering was ensured by \citet{svensson1980equity} using Szpilrajn's (\citeyear{Szpilrajn1930}) lemma.  
    These orderings do not satisfy \textit{continuity} (cf. the table in p.788 in \citealp{FM2003JME}). 

    \item \textit{Dropping One-Generation Additivity}: the  social welfare ordering $\succsim$ represented by the function $W: \mathcal{D} \to \mathbb{R}$ such that $W(\mathbf{u}) = \inf_{t\in\mathbb{N}} u_t$ for all $\mathbf{u} \in \mathcal{D}$.

    \item \textit{Dropping Mean Consistency}: the social welfare ordering $\succsim$ represented by the function $W: \mathcal{D} \to \mathbb{R}$ such that $W(\mathbf{u}) = \liminf_{t\rightarrow \infty} u_t$ for all $\mathbf{u} \in \mathcal{D}$.
\end{itemize}

Then, we verify the independence of the axioms in Theorem \ref{thm_chara_lCes}(2) and Theorem \ref{thm_genCesfixed}.

\begin{itemize}
    \item  \textit{Dropping Uniform Pareto}: the social welfare ordering $\succsim$ defined as for all $\mathbf{u}, \mathbf{v} \in \mathcal{D}$, $\mathbf{u} \sim \mathbf{v}$.

    \item \textit{Dropping Periodic Anonymity}: the  social welfare ordering $\succsim$ represented by the function $W: \mathcal{D} \to \mathbb{R}$ such that $W(\mathbf{u}) = u_1$ for all $\mathbf{u} \in \mathcal{D}$.

    \item \textit{Dropping Continuity}: the social welfare ordering $\succsim$ defined as for all $ \mathbf{u},  \mathbf{v} \in \mathcal{D}$, 
    \begin{align*}
        \bigg[ \exists T^\ast ~~ \text{s.t.} ~~ \forall T\geq T^\ast, ~~ \sum_{t = 1}^T u_t \geq \sum_{t = 1}^T v_t \bigg] \implies \mathbf{u} \succsim \mathbf{v}, 
        \\
        \bigg[ \exists T^\ast ~~ \text{s.t.} ~~ \forall T\geq T^\ast, ~~ \sum_{t = 1}^T u_t > \sum_{t = 1}^T v_t \bigg] \implies \mathbf{u} \succ \mathbf{v}. 
    \end{align*}
    The existence of such a social welfare ordering can be ensured by Szpilrajn's (\citeyear{Szpilrajn1930}) lemma, as in \citet{svensson1980equity}. These orderings do not satisfy \textit{continuity} (cf. the table in p.788 in \citealp{FM2003JME}). 

    \item \textit{Dropping Periodic Additivity}: the  social welfare ordering $\succsim$ represented by the function $W: \mathcal{D} \to \mathbb{R}$ such that $W(\mathbf{u}) = \inf_{t\in\mathbb{N}} u_t$ for all $\mathbf{u} \in \mathcal{D}$.

    \item \textit{Dropping Fixed-Step Replication Consistency}: the social welfare function  $\succsim$ represented by the function $W: \mathcal{D} \to \mathbb{R}$ such that for all $\mathbf{u} \in \mathcal{D}$, $W(\mathbf{u}) = \sup_{\mathbf{v} \in \mathrm{PS} (\mathbf{u})} \mu_\infty (\mathbf{v})$, where $\mathrm{PS} (\mathbf{u}) = \{ \mathbf{v} \in\mathcal{D}  \mid \text{$\mathbf{v}$ is a periodic stream such that $\mathbf{u} \geq \mathbf{v}$}\}$.
\end{itemize}

\section{Existence of characterized social welfare orderings}
\label{app_exist}

This section verifies the existence of the social welfare orderings characterized in Theorems \ref{thm_genCes1}--\ref{thm_full2}. 
First, we examine the social welfare orderings obtained in Theorem \ref{thm_full2}.  

\begin{observation}
\label{obs_exisfix}
    There exists a social welfare ordering characterized in Theorem \ref{thm_full2}. 
\end{observation}

\begin{proof}
    Note that by Observation \ref{obs_proLces}, $\ell^\text{Ces}$ is a linear subspace of $\ell^\infty$.
     Let $W: \ell^\text{Ces} \rightarrow \mathbb{R}$ be a linear function such that for all $\mathbf{u}\in\ell^\text{Ces}$, $W(\mathbf{u}) = \mu_\infty (\mathbf{u})$. 
    Consider the function $G:\ell^\infty \rightarrow \mathbb{R}$ defined as for all $\mathbf{u}\in \ell^\infty$, $G(\mathbf{u}) = \inf_{k\in \mathbb{N}}\limsup_{T\rightarrow \infty} \mu_{kT} (\mathbf{u})$.

    Note that for all $\mathbf{u}\in\ell^\text{Ces}$, $G(\mathbf{u}) = W(\mathbf{u})$ and $G$ is a convex function. 
    To see this, let $\mathbf{u}, \mathbf{v} \in \ell^\infty$ and $\alpha\in (0,1)$. 
    For all $\mathbf{w}\in \ell^\infty$, since the function $k \mapsto  \limsup_{T\rightarrow \infty} \mu_{kT} (\mathbf{w})$ on $\mathbb{N}$ is a lower-bounded, non-increasing function, we have
    \begin{equation*}
        G(\mathbf{w}) = \inf_{k\in \mathbb{N}}\limsup_{T\rightarrow \infty} \mu_{kT} (\mathbf{w}) = \lim_{k\rightarrow \infty} \limsup_{T\rightarrow \infty} \mu_{kT} (\mathbf{w}). 
    \end{equation*} 
    Therefore, by the linearity of the limit operations, 
    \begin{align*}
        G(\alpha \mathbf{u} + (1-\alpha) \mathbf{v}) 
        &= \lim_{k\rightarrow \infty} \limsup_{T\rightarrow \infty} \mu_{kT} (\alpha \mathbf{u} + (1-\alpha) \mathbf{v}) \\
        &\leq \lim_{k\rightarrow \infty} \bigg[ \alpha \limsup_{T\rightarrow \infty} \mu_{kT} ( \mathbf{u} )+ (1-\alpha) \limsup_{T\rightarrow \infty} \mu_{kT} ( \mathbf{v} ) \bigg] \\
        &=   \alpha \lim_{k\rightarrow \infty} \limsup_{T\rightarrow \infty} \mu_{kT} ( \mathbf{u} )+ (1-\alpha) \lim_{k\rightarrow \infty} \limsup_{T\rightarrow \infty} \mu_{kT} ( \mathbf{v} ) \\
        &= \alpha  G(\mathbf{u}) + (1-\alpha)G( \mathbf{v}), 
    \end{align*} 
    that is, $G$ is a convex function. 
     By the Hahn--Banach extension theorem (e.g. Theorem 5.53 in \citealp{AB2006Math}), there exists a linear function $\widetilde{W}:\ell^\infty\rightarrow \mathbb{R}$ such that for all $\mathbf{u}\in\ell^\text{Ces}$, $\widetilde{W} (\mathbf{u} ) = G(\mathbf{u}) = W (\mathbf{u} )$ and for all $\mathbf{v}\in \ell^\infty$, $G(\mathbf{v}) \geq \widetilde{W}(\mathbf{v})$. 

     Finally we prove that for all $\mathbf{u}\in \ell^\infty$,  $\widetilde{W} ( \mathbf{u}) \geq \sup_{k\in \mathbb{N}} \liminf_{T \rightarrow \infty}  \mu_{kT} (\mathbf{u}) $.  
     Suppose to the contrary that  $\widetilde{W} ( \mathbf{u})  < \sup_{k\in \mathbb{N}} \liminf_{T \rightarrow \infty}  \mu_{kT} (\mathbf{u}) $  for some $\mathbf{u} \in \ell^\infty$. 
     Then, we have $\inf_{k\in\mathbb{N}} \limsup_{T \rightarrow \infty} \mu_{kT} (\widetilde{W} (\mathbf{u}) \mathbf{1}_\mathbb{N} - \mathbf{u}) < 0$.
     By $\widetilde{W} (\mathbf{v} )\leq G(\mathbf{v}) $ for all $\mathbf{v}\in \ell^\infty$, 
    \begin{equation*}
        \widetilde{W} (\widetilde{W} (\mathbf{u}) \mathbf{1}_\mathbb{N} - \mathbf{u} )\leq G(\widetilde{W} (\mathbf{u}) \mathbf{1}_\mathbb{N} - \mathbf{u}) < 0.  
    \end{equation*}
    By the linearity and $\widetilde{W}(\mathbf{v}) = \mu_\infty (\mathbf{v} ) $ for all $\mathbf{v}\in \ell^\text{Ces}$, $\widetilde{W} (\mathbf{u}) < \widetilde{W}(\mathbf{u})$, a contradiction. 
\end{proof}

Since for all $\mathbf{u}\in \ell^\infty$, 
    \begin{align*}
        \liminf_{T \rightarrow \infty}  \mu_T (\mathbf{u}) \leq \sup_{k\in\mathbb{N}} \liminf_{T \rightarrow \infty}  \mu_{kT} (\mathbf{u}) 
        ~~~~~ \text{and} ~~~~~
        \inf_{k\in\mathbb{N}} \limsup_{T \rightarrow \infty} \mu_{kT}(\mathbf{u}) \leq \limsup_{T \rightarrow \infty}  \mu_T (\mathbf{u}), 
    \end{align*}
the following holds. 

\begin{observation}
    There exists a social welfare ordering characterized in Theorem \ref{thm_full1}. 
\end{observation}

By these two observations, social welfare orderings characterized in Theorems \ref{thm_genCes1} and \ref{thm_genCesfixed} also exist. 
However, note that the proof of Observation \ref{obs_exisfix} depends on the Hahn--Banach extension theorem, which is an implication of the Axiom of Choice or other non-constructive objects such as ultrafilters. 
Thus, these social welfare orderings may not be  constructable. 
For Theorems \ref{thm_genCes1} and \ref{thm_genCesfixed}, we verify the existence of orderings without relying on the Hahn--Banach extension theorem. 
We prove that $W^1$, $W^2$, $W^3$ and $W^4$ defined in Remark \ref{rem_exis} satisfy all of the  properties in Theorems \ref{thm_genCes1} and \ref{thm_genCesfixed}.

\begin{proof}[Proof that $W^1$ satisfies all properties.]
    It is straightforward to prove that $W^1$ is a continuous Cesàro average function.
    It is sufficient to prove that it is tail-monotone and respects $\succsim^{\text{fix}-C}$ because these properties immediately that the function is weakly monotone and respects $\succsim^{C}$. 
    
    `Tail-Monotonicity.' 
    Consider $  \mathbf{u},  \mathbf{v} \in \ell^\infty$ such that there exist  $s \in \mathbb{N}$ and $\varepsilon > 0$ such that $u_t \geq v_t + \varepsilon$ for all $t \geq s$. By the definition of $W^1$, we have
    \begin{align*}
        W^1 ( \mathbf{u}) - W^1 ( \mathbf{v}) &= \sup_{k\in\mathbb{N}} \liminf_{T \rightarrow \infty} \frac{1}{kT} \sum_{i = 1}^{kT}  u_i -  \sup_{k\in\mathbb{N}} \liminf_{T \rightarrow \infty}  \frac{1}{kT} \sum_{i = 1}^{kT}  v_i  \\
        &= \lim_{k\rightarrow \infty} \lim_{T \rightarrow \infty} \{ \inf_{t > T} \frac{1}{kt} \sum_{i = 1}^{kt}  u_i -  \inf_{t > T} \frac{1}{kt} \sum_{i = 1}^{kt}  v_i \} \\
        &=  \lim_{k\rightarrow \infty} \lim_{T \rightarrow \infty} \{ \inf_{t > T} \frac{1}{kt} \sum_{i = 1}^{kt} ( u_i - v_i + v_i) -  \inf_{t > T} \frac{1}{kt} \sum_{i = 1}^{kt}  v_i \} .
    \end{align*}
    Since  $T$ goes to infinity, it is sufficient to consider the case  where $kT > s$. Then, 
    \begin{align*}
        W^1 ( \mathbf{u}) - W^1 ( \mathbf{v} ) &\geq
        \lim_{k\rightarrow \infty} \lim_{T \rightarrow \infty} \{ \inf_{t > T} \frac{1}{kt} \sum_{i = 1}^{s} ( u_i - v_i) + \inf_{t > T} \frac{1}{kt} \sum_{i = s}^{kt} (u_i - v_i)  \\
        & ~~~~~~~~~~~~~~~ + \inf_{t > T} \frac{1}{kt} \sum_{i = 1}^{kt}  v_i -  \inf_{t > T} \frac{1}{kt} \sum_{i = 1}^{kt}  v_i \}  \\
        &\geq \lim_{k\rightarrow \infty} \lim_{T \rightarrow \infty}\inf_{t > T} \frac{(kt - s + 1) \varepsilon}{kt} \\
        &= \varepsilon. 
    \end{align*}

    `Respecting $\succsim^{\text{fix}-C}$.'  Note that 
    \begin{align*}
        W^1 ( \mathbf{u}) - W^1 ( \mathbf{v}) &=\lim_{k\rightarrow \infty} \lim_{T \rightarrow \infty} \{ \inf_{t > T} \frac{1}{kt} \sum_{i = 1}^{kt} ( u_i - v_i + v_i) -  \inf_{t > T} \frac{1}{kt} \sum_{i = 1}^{kt}  v_i \} \\
        &\geq \lim_{k\rightarrow \infty} \lim_{T \rightarrow \infty} \{ \inf_{t > T} \frac{1}{kt} \sum_{i = 1}^{kt} ( u_i - v_i )  + \inf_{t > T} \frac{1}{kt} \sum_{i = 1}^{kt}  v_i -  \inf_{t > T} \frac{1}{kt} \sum_{i = 1}^{kt}  v_i \}  \\
        &= \lim_{k\rightarrow \infty} \liminf_{T \rightarrow \infty} \frac{1}{kT} \sum_{i = 1}^{kT} ( u_i - v_i ). 
    \end{align*}
    Suppose that $ \mathbf{u} \succsim^{\text{fix}-C}  \mathbf{v}$, i.e., there exists $k^\ast \in \mathbb{N}$ such that for all $T \in \mathbb{N}$, $ \sum_{t = 1}^{k^\ast T} u_t \geq \sum_{t = 1}^{k^\ast T} v_t$. 
    Then,  we have $\liminf_{T \rightarrow \infty} \frac{1}{k^\ast T} \sum_{i = 1}^{k^\ast T} ( u_i - v_i ) \geq 0$. This implies that 
    \begin{equation}
    \label{eq_w1_ineq}
        \lim_{k\rightarrow \infty} \liminf_{T \rightarrow \infty} \frac{1}{kT} \sum_{i = 1}^{kT} ( u_i - v_i ) = \sup_{k \in \mathbb{N}} \liminf_{T \rightarrow \infty} \frac{1}{kT} \sum_{i = 1}^{kT} ( u_i - v_i )
        \geq
        \liminf_{T \rightarrow \infty} \frac{1}{k^\ast T} \sum_{i = 1}^{k^\ast T} ( u_i - v_i )
        \geq 0. 
    \end{equation}
    Therefore, we have $W^1 ( \mathbf{u}) > W^1 ( \mathbf{v})$, as required. 
\end{proof}

\begin{proof}[Proof that $W^2$ satisfies all properties. ]
    Similarly, it is straightforward to prove that  $W^2$ is a continuous Cesàro average function. 
    It is sufficient to prove that it is tail-monotone and respects $\succsim^{\text{fix}-C}$
    because these properties immediately that the function is weakly monotone and respects $\succsim^{C}$. 
    
    `Tail-Monotonicity.' Take arbitrary utility streams  $\mathbf{u},  \mathbf{v} \in \ell^\infty$ such that there exist  $s \in \mathbb{N}$ and $\varepsilon > 0$ such that $u_t \geq v_t + \varepsilon$ for all $t \geq s$.
    Then, we have
    \begin{align*}
        W^2 ( \mathbf{u})  -  W^2 ( \mathbf{v}) 
        &= 
        \lim_{\beta\rightarrow 1^-} \{ \inf_{\delta \in (\beta, 1)} (1- \delta) \sum_{t = 1}^{\infty} \delta^{t-1} u_t - \inf_{\delta \in (\beta, 1)} (1- \delta) \sum_{t = 1}^{\infty} \delta^{t-1} v_t \} \\
        &= 
        \lim_{\beta\rightarrow 1^-} \{ \inf_{\delta \in (\beta, 1)} (1- \delta) \sum_{t = 1}^{\infty} \delta^{t-1} (u_t - v_t + v_t) - \inf_{\delta \in (\beta, 1)} (1- \delta) \sum_{t = 1}^{\infty} \delta^{t-1} v_t \} \\
        &\geq \lim_{\beta\rightarrow 1^-} \{ \inf_{\delta \in (\beta, 1)} (1- \delta) \sum_{t = 1}^{s} \delta^{t-1} (u_t - v_t)  + \inf_{\delta \in (\beta, 1)} (1- \delta) \sum_{t = s+1}^{\infty} \delta^{t-1} (u_t - v_t)   \\*
        & ~~~~~~~~~~~ + \inf_{\delta \in (\beta, 1)} (1- \delta) \sum_{t = 1}^{\infty} \delta^{t-1} v_t  - \inf_{\delta \in (\beta, 1)} (1- \delta) \sum_{t = 1}^{\infty} \delta^{t-1} v_t \} \\
        &= \lim_{\beta\rightarrow 1^-} \{ \inf_{\delta \in (\beta, 1)} (1- \delta) \sum_{t = 1}^{s} \delta^{t-1} (u_t - v_t)  + \inf_{\delta \in (\beta, 1)} (1- \delta) \sum_{t = s+1}^{\infty} \delta^{t-1} \varepsilon \} \\
        &\geq \liminf_{\delta\rightarrow 1^-}  \delta^s \varepsilon \\
        &= \varepsilon. 
    \end{align*}
    Thus, we have $ W^2 ( \mathbf{u})  >   W^2 ( \mathbf{v}) $.

    `Respecting $\succsim^{\text{fix}-C}$.'
    In the same way as $W^1$, we can show 
    \begin{equation*}
        W^2 ( \mathbf{u}) - W^2 ( \mathbf{v}) 
        \geq \liminf_{\delta\rightarrow 1^-} (1- \delta) \sum_{t = 1}^{\infty} \delta^{t-1} (u_t - v_t). 
    \end{equation*}
    By Observation \ref{obs_duCes} and \eqref{eq_w1_ineq}, we have 
    \begin{equation*}
        W^2 ( \mathbf{u}) - W^2 ( \mathbf{v})  \geq \sup_{k \in \mathbb{N}} \liminf_{T \rightarrow \infty} \frac{1}{kT} \sum_{t = 1}^{kT} ( u_t - v_t ) \geq 0. \qedhere
    \end{equation*}
\end{proof}

In the same way, we can prove that $W^3$ and $W^4$ satisfy these properties. 

\bibliographystyle{econ}
\bibliography{reference}

\end{document}